\documentclass[fleqn,usenatbib]{mnras}

\usepackage{newtxtext,newtxmath}
\usepackage[T1]{fontenc}

\DeclareRobustCommand{\VAN}[3]{#2}
\let\VANthebibliography\thebibliography
\def\thebibliography{\DeclareRobustCommand{\VAN}[3]{##3}\VANthebibliography}

\usepackage{graphicx}	% Including figure files
\usepackage{amsmath}	% Advanced maths commands
\title[Luminous planets not trapped at pressure bumps]{Accreting luminous low-mass planets escape from migration traps at pressure bumps}
\author[O. Chrenko and R. O. Chametla]{
Ond\v{r}ej Chrenko$^{1}$\thanks{E-mail: chrenko@sirrah.troja.mff.cuni.cz}
and Ra\'{u}l O. Chametla$^{1}$
\\
$^{1}$Charles University, Fac Math \& Phys, Astronomical Institute, V Hole\v{s}ovi\v{c}k\'{a}ch 747/2, 180 00 Prague 8, Czech Republic%\\
}

\date{Accepted 2023 July 5. Received 2023 June 21; in original form 2023 April 25}

\pubyear{2023}

\begin{document}
\label{firstpage}
\pagerange{\pageref{firstpage}--\pageref{lastpage}}
\maketitle

% Abstract of the paper
\begin{abstract}

We investigate the migration of Mars- to super-Earth-sized planets in the vicinity of a pressure bump in a 3D radiative protoplanetary disc while accounting for the effect of accretion heat release. 
Pressure bumps have often been assumed to act as efficient migration traps, but we show that the situation changes when the thermal forces are taken into account.
Our simulations reveal that for planetary masses $\lesssim$$2\,M_{\earth}$, once their luminosity exceeds the critical value predicted by linear theory, thermal driving causes their orbits to become eccentric, quenching the positive corotation torque responsible for the migration trap. As a result, planets continue migrating inwards past the pressure bump. Additionally,
we find that planets that remain circular and evolve in the super-Keplerian region of the bump exhibit a reversed asymmetry of their thermal lobes, with the heating torque having an opposite (negative) sign compared to the standard circular case, thus leading to inward migration as well.
We also demonstrate that the super-critical luminosities of planets in question can be reached through the accretion of pebbles accumulating in the bump. Our findings have implications for planet formation scenarios that rely on the existence of migration traps at pressure bumps, as the bumps may repeatedly spawn inward-migrating low-mass embryos rather than harbouring newborn planets until they become massive.

\end{abstract}

% Select between one and six entries from the list of approved keywords.
% Don't make up new ones.

%%%%%%%%%%%%%%%%%%%%%%%%%%%%%%%%%%%%%%%%%%%%%%%%%%

%%%%%%%%%%%%%%%%% BODY OF PAPER %%%%%%%%%%%%%%%%%%

\section{Introduction}
\label{sec:intro}

The recent advancements of interferometric observations
with high angular resolution have enabled to detect
ring-like concentrations of small solid particles
(dust or pebbles)
in numerous protoplanetary discs
\citep[e.g.][]{Andrews_etal_2018ApJ...869L..41A,Dipierro_etal_2018MNRAS.475.5296D,Cieza_etal_2019MNRAS.482..698C}.
Such ring-shaped accumulations could have formed at the locations of pressure bumps
where the disc rotation becomes locally super-Keplerian and the inward drift
of small solids due to the aerodynamic drag is blocked \citep{Nakagawa_etal_1986Icar...67..375N,Dullemond_etal_2018ApJ...869L..46D,Teague_etal_2018ApJ...868..113T}.

Since pressure bumps overlap with transitions in the gas density,
they can act as barriers for planetary migration due to a local boost
of the positive corotation torque, which balances the negative Lindblad torque \citep[e.g.][]{Masset_etal_2006ApJ...642..478M,Ataiee_Kley_2021A&A...648A..69A,Chrenko_etal_2022A&A...666A..63C}. 
Therefore, a pressure bump could hypothetically represent a sweat spot for planet formation:
Any planet growing at the bump would be protected from planetary migration and 
it would remain submerged in a relatively dense and self-replenishing 
reservoir of solids from which it could continue accreting.
Such an interplay has become a key component of many novel planet
formation scenarios \citep{Morbidelli_2020A&A...638A...1M,Guilera_etal_2020A&A...642A.140G,Chambers_2021ApJ...914..102C,Andama_etal_2022MNRAS.512.5278A,Lau_etal_2022A&A...668A.170L,Jiang_Ormel_2023MNRAS.518.3877J}.

However, the existence of the migration trap at the pressure bump 
has so far been justified by taking only the Lindblad and corotation
disc-driven torques \citep{Goldreich_Tremaine_1979ApJ...233..857G,Ward_1991LPI....22.1463W,Masset_2001ApJ...558..453M,Tanaka_etal_2002ApJ...565.1257T} into account
and thus, from the viewpoint of planet-disc interactions,
the picture is not entirely complete.
In non-isothermal discs with any form of thermal diffusion,
planets are subject to additional thermal torques \citep{Masset_2017MNRAS.472.4204M}.
When the planet is non-luminous, the gas traveling past the planet gains energy
by compressional heating but this energy is spread by thermal diffusion
\citep{Lega_etal_2014MNRAS.440..683L} and a perturbation arises which
is cooler and denser compared to a fully adiabatic case.
When the planet is accreting and luminous, a threshold luminosity exists
for which the disc perturbation is similar to the fully adiabatic case \citep{Masset_2017MNRAS.472.4204M}.
Planets with super-critical luminosities
switch to the regime of the heating torque \citep{Benitez-Llambay_etal_2015Natur.520...63B,Hankla_etal_2020ApJ...902...50H,Chametla_Masset_2021MNRAS.501...24C}.

The heating torque arises because the accretion heat renders the gas flowing
past the planet underdense. The underdense gas is redistributed by
the thermal (or radiative) diffusion and disc shear
and two lobes are formed,
the inner lobe leading and the outer lobe trailing the orbital motion of the planet \citep{Benitez-Llambay_etal_2015Natur.520...63B,Masset_2017MNRAS.472.4204M}.
For a typical sub-Keplerian disc and a circular planetary orbit, the outer
trailing lobe is more pronounced because the corotation radius between the planet
and the disc material is shifted inwards.
We recall, however, that
\cite{Chrenko_Lambrechts_2019} showed that the advection of hot gas near 
super-Earth-sized planets
is not governed purely by the shear motion but rather by a complex 3D circumplanetary flow
interacting with the horseshoe region of the planet \citep[see also][]{Chametla_Masset_2021MNRAS.501...24C}.

Additionally, the perturbing force acting upon the luminous planet can also excite 
its orbital eccentricity \citep{Chrenko_etal_2017A&A...606A.114C,Eklund_Masset_2017MNRAS.469..206E,Fromentau_Masset_2019MNRAS.485.5035F,Velasco-Romero_etal_2022MNRAS.509.5622V,Cornejo_etal_2023MNRAS.tmp..658C}.
When that happens, the two thermal lobes are replaced with a single lobe 
whose trajectory can be well approximated with an epicycle
and the planet is said to enter the headwind-dominated regime of thermal torques \citep{Eklund_Masset_2017MNRAS.469..206E}.
While the heating torque is typically positive and supports outward migration
in the circular shear-dominated
regime \citep{Benitez-Llambay_etal_2015Natur.520...63B,Masset_2017MNRAS.472.4204M},
its contribution to the overall torque balance in the eccentric headwind-dominated regime 
is less clear.

In our study, we investigate the migration of pebble-accreting planets
in the vicinity of the pressure bump while taking the thermal torques into account,
with the aim to answer the following questions.
Will the thermal torques assist or counteract the migration trap? Will the orbital 
eccentricity grow and if so, how will the torque balance change?
By carrying out an extensive set of numerical simulations, we show 
that planets with super-critical luminosities are likely to experience the 
eccentricity excitation and enter the headwind-dominated regime.
The subsequent orbital migration of such planets is
dominated by the Lindblad torque because the corotation torque is quenched
for eccentric orbits \citep{Bitsch_Kley_2010A&A...523A..30B,Fendyke_Nelson_2014MNRAS.437...96F}
and we demonstrate that the influence of the thermal torque 
on the evolution of the semi-major axis weakens as well \citep[see also][]{Pierens_2023MNRAS.520.3286P}.
Finally, for parameters that allow the planet to remain
in the circular shear-dominated regime, we show that the thermal torques
in the super-Keplerian region of the pressure bump have an opposite effect
compared to \cite{Benitez-Llambay_etal_2015Natur.520...63B}.

%The article is structured as follows.

\section{3D radiative hydrodynamic model}
\label{sec:model}

We model a patch of a 3D gas disc on an
Eulerian grid with uniform
spacing and $N_{r}\times N_{\theta} \times N_{\phi}$
cells, where the subscripts stand for the radial, azimuthal,
and colatitudinal spherical coordinates, respectively.
The azimuthal extent of the grid 
for our simulations with planets covers a quadrant,
not the full azimuth (see Section~\ref{sec:inicond}).
We use the Fargo3D code
\citep{Benitez-Llambay_Masset_2016ApJS..223...11B}
and treat the gas disc as a viscous fluid evolving in
a non-inertial frame centered on the star $M_{\star}$ and co-rotating with an embedded planet $M_{\mathrm{p}}$,
whose orbital evolution we study. The gravitational
potential of the planet is smoothed with the cubic spline
function of \cite{Klahr_Kley_2006A&A...445..747K} at cell-planet distances $d<r_{\mathrm{sm}}$, where $r_{\mathrm{sm}}$
is the smoothing length (Table~\ref{tab:param}), as
\begin{equation}
    \Phi_{\mathrm{p}} = - \frac{GM_{\mathrm{p}}}{d}\left[\left(\frac{d}{r_{\mathrm{sm}}}\right)^{4}-2\left(\frac{d}{r_{\mathrm{sm}}}\right)^{3}+2\frac{d}{r_{\mathrm{sm}}}\right] \, ,
\end{equation}
where $G$ is the gravitational constant.

Aside from the continuity and momentum equations that
are included in the public version of Fargo3D \citep{Benitez-Llambay_Masset_2016ApJS..223...11B},
we consider the evolution of
the internal energy density of the gas $\epsilon$
and the energy density of diffuse thermal radiation $E_{\mathrm{R}}$ following the two-temperature approximation
\citep{Commercon_etal_2011A&A...529A..35C,Bitsch_etal_2013A&A...549A.124B}:
\begin{equation}
\frac{\partial E_{\mathrm{R}}}{\partial t} + \nabla\cdot\vec{F} = \rho \kappa_{\mathrm{P}} \left[4\sigma T^{4} - cE_{\mathrm{R}}\right] \, ,
\label{eq:e_rad}
\end{equation}
\begin{equation}
    \frac{\partial \epsilon}{\partial t} + \nabla\cdot\left(\epsilon\vec{v}\right) = 
    - \rho\kappa_{\mathrm{P}}\left[4\sigma T^{4} - cE_{\mathrm{R}}\right]
    -P\nabla\cdot\vec{v}
    + Q_{\mathrm{visc}} + Q_{\mathrm{art}} + Q_{\mathrm{acc}} \, ,
    \label{eq:e_gas}
\end{equation}
where $t$ is the time, $\vec{F}$ is the radiation flux vector, $\rho$ is the gas density, $\kappa_{\mathrm{P}}$ is the Planck opacity, $\sigma$ is the Stefan-Boltzmann constant, $T$ is the gas temperature, $c$ is the speed
of light, $\vec{v}=(v_{r},v_{\theta},v_{\phi})$ is the velocity vector of the gas flow,
$P$ is the gas pressure ($P=(\gamma-1)\epsilon$ for the ideal gas with the
adiabatic index $\gamma$), 
$Q_{\mathrm{visc}}$ is the viscous heating term \citep{Mihalas_WeibelMihalas_1984frh..book.....M},
$Q_{\mathrm{art}}$ is the heating due to the shock-spreading
viscosity of finite-difference codes \citep{Stone_Norman_1992ApJS...80..753S},
and $Q_{\mathrm{acc}}$ is the heat source related to the luminosity of the embedded planet. For the implementation of the energy equations,
the flux-limited approximation, and the remaining closure relations, the reader is referred to \cite{Chrenko_Lambrechts_2019} and references therein.
Let us only point out that 
the Planck and Rosseland opacities (the latter of which governs the radiation diffusion) assumed here 
are uniform and equal, $\kappa_{\mathrm{P}}=\kappa_{\mathrm{R}}=\kappa$. 
Equations (\ref{eq:e_rad}) and (\ref{eq:e_gas}) are solved implicitly using the successive over-relaxation method with the relative precision of $10^{-8}$.

To model the heat release due to planetary accretion,
we consider a simple luminosity relation \citep{Benitez-Llambay_etal_2015Natur.520...63B}
\begin{equation}
    L = \frac{GM_{\mathrm{p}}\dot{M_{\mathrm{p}}}}{R_{\mathrm{p}}} = \frac{GM_{\mathrm{p}}^{2}}{R_{\mathrm{p}}\tau} \, ,
    \label{eq:L}
\end{equation}
where $\dot{M_{\mathrm{p}}}$ is the mass accretion rate of the planet, $R_{\mathrm{p}}$ is its physical radius, and
$\tau=M_{\mathrm{p}}/\dot{M_{\mathrm{p}}}$ is its mass doubling time.
Let us point out that $\tau$ is used
only to regulate $L$ but $M_{\mathrm{p}}$
itself is kept fixed throughout our simulations.
The heat source is non-zero in eight cells surrounding
the planet \citep{Benitez-Llambay_etal_2015Natur.520...63B}
and zero elsewhere. To account for the shift of the planet
with respect to these cells, we define the grid
coordinates of the planet
$(r_{\mathrm{p}},\theta_{\mathrm{p}},\phi_{\mathrm{p}})$,
coordinates of an $n$-th
cell centre $(r_{n},\theta_{n},\phi_{n})$,
dimensions of the respective cell
$(\Delta r,\Delta \theta,\Delta \phi)$, its volume $V_{n}$, and apply
\citep[e.g.][]{Velasco-Romero_etal_2022MNRAS.509.5622V}
\begin{equation}
    Q_{\mathrm{acc},n} = \frac{L}{V_{n}}\left(1-\frac{|r_{\mathrm{p}}-r_{n}|}{\Delta r}\right)\left(1-\frac{|\theta_{\mathrm{p}}-\theta_{n}|}{\Delta \theta}\right)\left(1-\frac{|\phi_{\mathrm{p}}-\phi_{n}|}{\Delta \phi}\right) \, ,
    \label{eq:q_acc}
\end{equation}
inside cells that satisfy $|r_{\mathrm{p}}-r_{n}|<\Delta r$,
$|\theta_{\mathrm{p}}-\theta_{n}| < \Delta \theta$,
and $|\phi_{\mathrm{p}}-\phi_{n}|<\Delta \phi$.

Most of our simulations are performed on a (i) disc quadrant
and with (ii) migrating planets. Due to (i), 
it is not possible to consider the indirect potential
term related to the acceleration of the star generated
by the disc. Point (ii) motivates us to subtract
the azimuthally-averaged gas density from $\rho$ before
the disc-planet interaction is evaluated \citep{Baruteau_Masset_2008ApJ...678..483B}. Such a
procedure improves the displacement of the Lindblad resonances in a non-self-gravitating disc.
The planetary orbit is propagated using the standard fourth-order
Runge-Kutta integrator of Fargo3D.

\begin{table}
	\centering
	\caption{Fixed parameters for simulations with embedded
 planets (simulations of disc relaxation differ in the grid span and resolution; Section~\ref{sec:inicond}).
 Below, $r_{\mathrm{p}}$ marks the initial radial planet position and $R_{\mathrm{H}}$ is the planet's Hill radius.}
	\label{tab:param}
	\begin{tabular}{lcc} % four columns, alignment for each
		\hline
        \hline
		Parameter & Notation & Value \\
     	\hline
      Grid resolution & $N_{r}\times N_{\theta}\times N_{\phi}$ & $1536\times4096\times128$ \\
      Radial grid span & $(r_{\mathrm{min}},r_{\mathrm{max}})$ & $(0.7,1.3)\,r_{\mathrm{p}}$ \\
      Azimuthal grid span & $(\theta_{\mathrm{min}},\theta_{\mathrm{max}})$ & $(-\pi/4,\pi/4)$ \\
      Vertical grid span & $(\phi_{\mathrm{min}},\phi_{\mathrm{max}})$ & $(7\pi/15,\pi/2)$ \\
      Stellar mass & $M_{\star}$ & $1\,M_{\sun}$ \\
      Surface density (at 1 au) & $\Sigma_{0}$ & $456\,\mathrm{g}\,\mathrm{cm}^{-2}$ \\
      Adiabatic index & $\gamma$ & $1.43$ \\
      Mean molecular weight & $\mu$ & $2.3$ \\
      Kinematic viscosity & $\nu$ & $10^{15}\,\mathrm{cm}^{2}\,\mathrm{s}^{-1}$ \\
      Opacity & $\kappa$ & $1\,\mathrm{cm}^{2}\,\mathrm{g}^{-1}$ \\
      Bump amplitude & $A_{\mathrm{b}}$ & $1.45$ \\
      Bump centre & $r_{\mathrm{b}}$ & $5.2\,\mathrm{au}$ \\
      Bump width & $w_{\mathrm{b}}$ & $0.32\,\mathrm{au}$ \\
      Smoothing length & $r_{\mathrm{sm}}$ & $0.15\,R_{\mathrm{H}}$ \\
		\hline
	\end{tabular}
\end{table}

\subsection{Initial conditions with a pressure bump}
\label{sec:inicond}

Before conducting simulations with embedded planets,
it is imperative to find an equilibrium state of the
unperturbed disc determined by the balance between
viscous heating and radiative cooling\footnote{Realistic
discs are also heated by stellar irradiation
\citep[e.g.][]{Bitsch_etal_2014A&A...564A.135B,Chrenko_Nesvorny_2020A&A...642A.219C}
but we neglect it here because the vertical opening angle of the domain is small (to reach high resolution), which prevents us from resolving the absorption of impinging stellar photons. Therefore, our model is applicable
to optically thick disc regions within several au rather
than to outer disc regions where stellar irradiation
dominates the energy budget.}.
We start with a disc that has a radial extension of
$(r_{\mathrm{min}},r_{\mathrm{max}})=(2.6,7.8)\,\mathrm{au}$
and a grid resolution of $N_{r}\times N_{\theta} \times N_{\phi} = 768 \times 1 \times 64$. The disc is azimuthally symmetric (as represented by $N_{\theta}=1$) as well as
vertically symmetric around the midplane (only one hemisphere of the disc is modelled). Remaining parameters
are the same as in Table~\ref{tab:param}.

The initial temperature
profile is that of an optically thin disc \citep[e.g.][]{Ueda_etal_2017ApJ...843...49U} and 
the initial surface
density profile follows
\begin{equation}
    \Sigma(r) = \Sigma_{0}\left(\frac{r}{1\,\mathrm{au}}\right)^{-0.5} \, .
    \label{eq:sigma}
\end{equation}
We introduce the pressure bump as a Gaussian perturbation of
the surface density \citep{Pinilla_etal_2012A&A...538A.114P,Dullemond_etal_2018ApJ...869L..46D}
\begin{equation}
    \Sigma'(r) = \Sigma(r)\left[1 + (A_{\mathrm{b}}-1)\exp{\left(-\frac{(r-r_{\mathrm{b}})^{2}}{2w_{\mathrm{b}}^{2}}\right)}\right] \, ,
\end{equation}
where $A_{\mathrm{b}}$ is the bump amplitude, $r_{\mathrm{b}}$ is the radial centre of the perturbation,
and $w_{\mathrm{b}}$ parametrizes the width of the Gaussian.
To minimize the viscous spreading over the course of our simulations, we `mirror' the Gaussian perturbation
as a minimum in the viscosity profile \citep[e.g.][]{Ataiee_etal_2014A&A...572A..61A}
\begin{equation}
    \nu'(r) = \nu\left[1 + (A_{\mathrm{b}}-1)\exp{\left(-\frac{(r-r_{\mathrm{b}})^{2}}{2w_{\mathrm{b}}^{2}}\right)}\right]^{-1} \, .
\end{equation}
Let us point out that in the absence of the pressure bump,
the disc would be similar to that of \cite{Eklund_Masset_2017MNRAS.469..206E}.

The first part of our relaxation procedure is hydrostatic
and iterative. For a given temperature field, we solve
the equations of hydrostatic equilibrium
\citep{Flock_etal_2016ApJ...827..144F,Chrenko_Nesvorny_2020A&A...642A.219C}
to find a new density distribution $\rho(r,\phi)$.
Then we fix the new density distribution and perform one
time step equal to the characteristic radiation diffusion time-scale in the energy equations. With the updated temperature field, the iteration process is repeated until
the relative change in $T$ and $\rho$ per iteration becomes as small as $10^{-5}$.

In the second step of our relaxation procedure, we continue
to evolve the disc using the full set of time-dependent
fluid equations over 2,000 orbital time-scales at $r_{\mathrm{b}}$. To insert a planet at an arbitrary position
$r_{\mathrm{p}}$ in the disc, we remap the relaxed disc to a numerical grid with the radial extension of 
$(r_{\mathrm{min}},r_{\mathrm{max}})=(0.7,1.3)\,r_{\mathrm{p}}$ and an azimuthal extension of a quadrant.
The resolution of the remapped grid is given in Table~\ref{tab:param}. We also transform the azimuthal
gas velocity $v_{\theta}$ to a frame corotating with the planet. If the planet starts in an eccentric orbit,
the frame velocity is corrected for the eccentricity
(planets start in their apocentre).

\subsection{Boundary conditions}

The boundaries are periodic in the azimuthal direction
and the disc is mirrored at the midplane.
Scalar quantities $(\rho,\epsilon,E_{\mathrm{R}})$ have symmetric boundary conditions,
with the exception of $E_{\mathrm{R}}$ at the
lower boundary in the colatitude, where 
we allow for the escape of photons through the disc surface by setting
$E_{\mathrm{R}}=a_{\mathrm{R}}T_{\mathrm{bc}}^{4}$, $a_{R}$ being 
the radiation constant and $T_{\mathrm{bc}}=5\,\mathrm{K}$.
Azimuthal velocities are symmetric at the vertical boundaries
and a Keplerian extrapolation is used at the radial boundaries.
Radial/vertical velocities are anti-symmetric at radial/vertical boundaries, respectively, and symmetric elsewhere.

Boundary conditions are supplemented with the wave-damping
zones \citep{deValBorro_etal_2006MNRAS.370..529D} in the
radial intervals of $(1,1.2^{2/3})r_{\mathrm{min}}$
and $(1.2^{-2/3},1)r_{\mathrm{max}}$. We damp $\rho$, $v_{r}$ and $v_{\phi}$ towards their values
corresponding to the end of the disc relaxation.
Azimuthal velocities and energy densities are not damped
and there is no
damping zone at the disc surface, nor in the midplane.

\section{Results}
\label{sec:results}

\begin{figure}
    \centering
    \includegraphics[width=0.95\columnwidth]{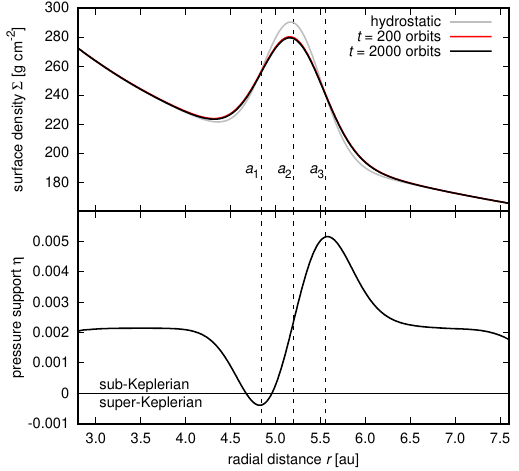}
    \caption{\emph{Top}:
    Radial profile of the gas surface density $\Sigma$ after the
    hydrostatic
    relaxation (grey curve), after 200 orbits of the hydrodynamic relaxation (red curve), and at the end of the hydrodynamic relaxation (black curve). \emph{Bottom}: Radial profile of the pressure support parameter $\eta$ in the relaxed disc.
    In \emph{both panels}, vertical dashed lines mark the semi-major axes of planets considered in this work, $a=4.84$, $5.2$,
    and $5.56\,\mathrm{au}$.
    The horizontal $\eta=0$ line in the \emph{bottom panel} 
    separates the regimes of sub-Keplerian ($\eta>0$) and super-Keplerian ($\eta<0$) orbital velocities of the gas.}
    \label{fig:profiles}
\end{figure}

\begin{figure}
    \centering
    \includegraphics[width=0.95\columnwidth]{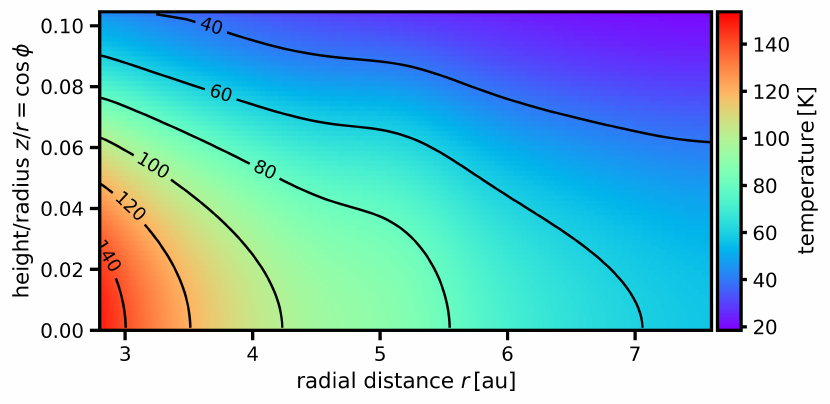}
    \caption{Gas temperature $T$ in the vertical plane of
    the relaxed disc before planet insertion. Black curves depict several temperature isocontours.}
    \label{fig:temper}
\end{figure}
\begin{figure}
    \centering
    \includegraphics[width=0.95\columnwidth]{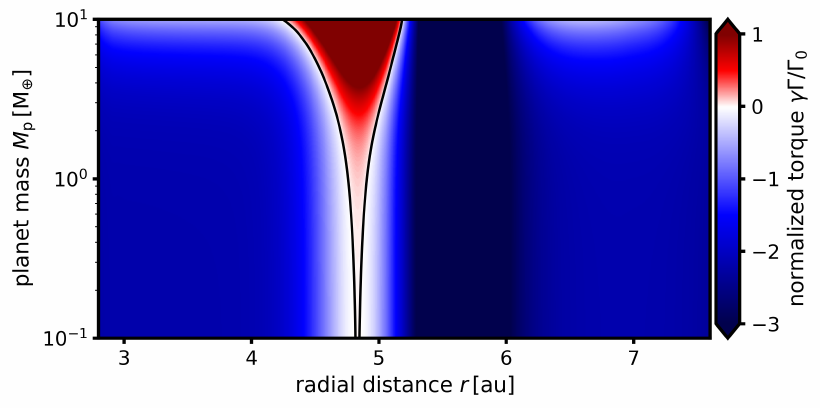}
    \caption{Estimate of the normalized gas-driven
    torque $\gamma\Gamma/\Gamma_{0}$ as a function of the planetary mass $M_{\mathrm{p}}$
    and radial distance $r$ based on the formulae of \protect\cite{Jimenez_Masset_2017MNRAS.471.4917J}, i.e. in the 
    absence of thermal torques. Black isocontours mark the zero-torque locations; planets are expected to become trapped at the outer black curve where the predicted migration is convergent.}
    \label{fig:tqmap}
\end{figure}

\subsection{Relaxed disc}

Fig.~\ref{fig:profiles} (top panel) shows the radial profiles
of the gas surface density $\Sigma$ obtained during the relaxation
towards the thermodynamic equilibrium. Starting from our hydrostatic
estimate, the bump amplitude slightly decreases and the bump edges
undergo minor spreading over the first 200 dynamical time-scales
of the hydrodynamic relaxation. Afterwards,
the red and black curves are difficult to distinguish 
from one another and the profile remains
almost unchanged.

Pressure bumps are typically characterized by the variation
of the pressure support that a gas parcel feels while orbiting
in the disc. We computed the pressure support parameter in the midplane \citep[e.g.][]{Nakagawa_etal_1986Icar...67..375N}
\begin{equation}
    \eta = - \frac{1}{2}\left(\frac{H}{r}\right)^{2}\frac{\partial\log P}{\partial\log r} \, ,
    \label{eq:eta}
\end{equation}
where $H$ is the non-isothermal pressure scale height that
relates to the sound speed $c_{\mathrm{s}}=\sqrt{\gamma P/\rho}$
and the local Keplerian frequency $\Omega_{\mathrm{K}}=\sqrt{GM_{\star}/r^{3}}$ as $H=c_{\mathrm{s}}/(\sqrt{\gamma} \Omega_{\mathrm{K}})$. If $\eta>0$, the gas parcel orbits
at a sub-Keplerian velocity. If, on the other hand, $\eta<0$,
the gas parcel becomes super-Keplerian. Fig.~\ref{fig:profiles}
(bottom panel)
indicates that $\eta\simeq0.002$ away from the pressure bump,
the maximum value of $\eta\simeq0.005$ is reached roughly
in the outer half of the surface density peak,
and the minimum value
of $\eta\simeq-4\times10^{-4}$ is reached in the inner half
of the $\Sigma$ peak.

In Fig.~\ref{fig:temper}, we show the gas temperature $T$
in the vertical plane of the disc.
The vertical temperature stratification is rather steep,
as expected for a disc dominated by the viscous heating,
and it is apparent that the bump does not have a strong influence on $T$.
The bump only slightly flattens the radial temperature gradient,
which results in the leveling of isocontours with the horizontal
axis near 5 au in Fig.~\ref{fig:temper}.

\subsection{Bump parameters and the migration trap}
\label{sec:bump_param}

The span of possible parameters characterizing pressure bumps
in protoplanetary discs is largely unconstrained because
the physics of these bumps is still a subject of intensive research.
We choose the bump position $r_{\mathrm{b}}=5.2\,\mathrm{au}$
to tie our study of planet migration to previous works \citep{Benitez-Llambay_etal_2015Natur.520...63B,Eklund_Masset_2017MNRAS.469..206E}.
The Gaussian width $w_{\mathrm{b}}=0.32\,\mathrm{au}\simeq1.5H$ is chosen to ensure that the bump remains Rossby-stable \citep[wider than $H$;][]{Lovelace_etal_1999ApJ...513..805L,Dullemond_etal_2018ApJ...869L..46D}.
Regarding the bump amplitude $A_{\mathrm{b}}=1.45$, it is kept rather
small but large enough to facilitate the existence of (i) a radial
interval of super-Keplerian gas rotation and (ii) a migration trap in
the absence of thermal torques, even for planets with very low masses.
Point (i) is fulfilled based on Fig.~\ref{fig:profiles} and point 
(ii) is proven in the following paragraph. As such, our bump can be interpreted
to be close to the lower limit of $A_{\mathrm{b}}$ and
$w_{\mathrm{b}}$ (while satisfying all of the above-mentioned
requirements). Larger values of $A_{\mathrm{b}}$ and
$w_{\mathrm{b}}$ are not ruled out by our study.

To verify that the pressure bump would act as a migration trap
in the absence of thermal torques, we applied the torque formulae
of \cite{Jimenez_Masset_2017MNRAS.471.4917J} to our disc model
and constructed the migration map shown in Fig.~\ref{fig:tqmap}.
According to the obtained result, low-mass planets would migrate
inwards due to the dominance of the Lindblad torque
in the majority of the disc \citep{Tanaka_etal_2002ApJ...565.1257T}.
However, there is indeed
a narrow interval of radii for which the disc torque would become
positive due to a boost of the corotation torque (red colour in Fig.~\ref{fig:tqmap}) and the planets would experience outward migration \citep{Masset_etal_2006ApJ...642..478M}.
At the outer edge of the red-coloured
region, there is the zero-torque radius that would act as a migration trap in the pressure bump.

Having analyzed the disc properties and their influence on
the Lindblad and corotation torques, we specify three values of
interest for the planetary semi-major axes $a$.
Our fiducial value is $a_{1}=4.84,\mathrm{au}$ and it is
marked in Fig.~\ref{fig:profiles} with the innermost vertical
dashed line. At this disc location, the planet starts within
the minimum of $\eta$ and at the same time, it finds itself
within the V-shaped region of outward migration in Fig.~\ref{fig:tqmap}.
The second semi-major axis of interest is $a_{2}=5.2\,\mathrm{au}$ (middle dashed line in Fig.~\ref{fig:profiles})
for which the planet is close to the maximum
of the surface density peak while $\eta$ is rather similar to the 
background unperturbed value.
Finally, we choose $a_{3}=5.56\,\mathrm{au}$ (outermost dashed line in Fig.~\ref{fig:profiles}) as a counterpart to $a_{1}=4.84\,\mathrm{au}$ because it overlaps
with the maximum of $\eta$.

\subsection{Characteristic scales}

The linear perturbation theory of thermal torques 
\citep{Masset_2017MNRAS.472.4204M} argues that the thermal
disturbance near a luminous planet has a characteristic length-scale
\begin{equation}
    \lambda_{\mathrm{c}}=\sqrt{\frac{\chi}{(3/2)\Omega_{\mathrm{K}}\gamma}} \, ,
    \label{eq:lambda}
\end{equation}
where $\chi$ is the thermal diffusivity. Recent high-resolution simulations of 
inviscid discs with thermal diffusivity \citep{Chametla_Masset_2021MNRAS.501...24C,Velasco-Romero_etal_2022MNRAS.509.5622V}
have demonstrated that the numerical convergence of thermal forces depends
on the resolution and requires at least 10 cells per $\lambda_{\mathrm{c}}$, or
$l/\lambda_{\mathrm{c}}=0.1$ where $l$ is the cell size.

In order to verify that our resolution is sufficient, we calculate the thermal diffusivity
due to radiation diffusion in an optically thick medium as \citep{Kley_etal_2009A&A...506..971K,Bitsch_Kley_2011A&A...536A..77B}
\begin{equation}
    \chi = \frac{16\gamma(\gamma-1)\sigma T^{4}}{3\kappa(\rho H \Omega_{\mathrm{K}})^{2}} \, ,
    \label{eq:chi}
\end{equation}
and, at $a_{1}=4.84\,\mathrm{au}$, we obtain $\chi=3.82\times10^{15}\,\mathrm{cm}^{2}\,\mathrm{s}^{-1}$ and $\lambda_{\mathrm{c}}=0.02\,\mathrm{au}=0.1H$.
Using the quadrant grid in the azimuth and the resolution specified in Table~\ref{tab:param}, we reach $l_{r}/\lambda_{\mathrm{c}}=0.09$, $l_{\theta}/\lambda_{\mathrm{c}}=0.09$, and $l_{\phi}/\lambda_{\mathrm{c}}=0.2$, where $l_{r,\theta,\phi}$ are the lengths of cell interfaces along the respective spherical coordinates. Therefore,
the thermal disturbance is well resolved in the radial and azimuthal directions
and only slightly under-resolved in the vertical direction, which we consider
a reasonable compromise.

Additionally, Type I migration critically depends on the resolution of the half-width of the horseshoe
region $x_{s}$ \citep{Paardekooper_etal_2011MNRAS.410..293P,Jimenez_Masset_2017MNRAS.471.4917J},
which influences the accuracy of the horseshoe drag \citep{Ward_1991LPI....22.1463W,Masset_2001ApJ...558..453M,Baruteau_Masset_2013LNP...861..201B}. The minimal requirement for radiative discs is to resolve $x_{s}$ by 4 cells \citep{Lega_etal_2014MNRAS.440..683L}.
For the lowest planetary mass that we shall consider in the following (see Table~\ref{tab:paramspace} for the range of parameters varied in our simulations), that is $M_{\mathrm{p}}=0.1\,M_{\earth}$, we have $(7,7,3)$ cells per $x_{s}$
in the $(r,\theta,\phi)$ directions. For $M_{\mathrm{p}}=1\,M_{\earth}$,
we have $(21,21,10)$ cells per $x_{s}$, respectively.
For completeness, let us remark the resolution of the Hill radius $R_{\mathrm{H}}$; we have $(12,12,6)$ cells per $R_{\mathrm{H}}$
for the Mars-mass planet and $(26,26,12)$ cells per $R_{\mathrm{H}}$ for
the Earth-mass planet.

\begin{figure*}
    \centering
    \begin{tabular}{cc}
    \includegraphics[width=0.95\columnwidth]{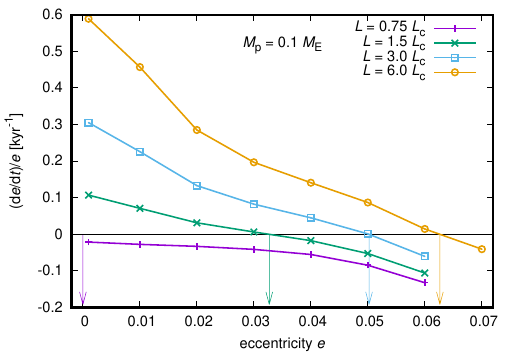} &
    \includegraphics[width=0.95\columnwidth]{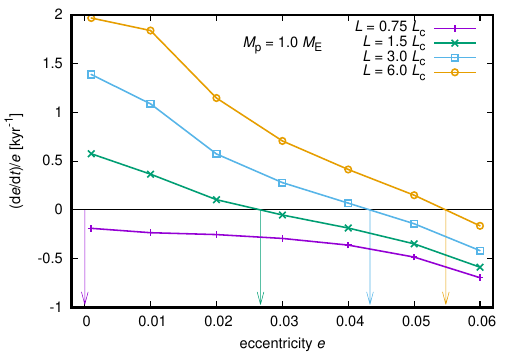} \\
    \includegraphics[width=0.95\columnwidth]{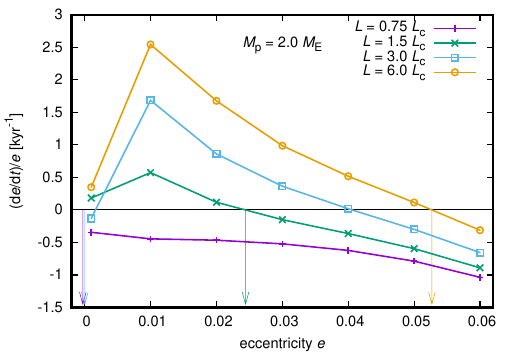} & 
    \includegraphics[width=0.95\columnwidth]{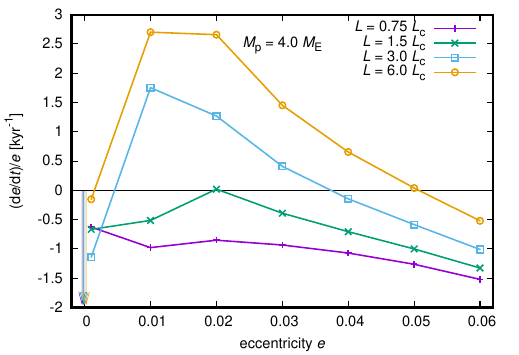} \\
    \end{tabular}
    \caption{Eccentricity evolution rate $\dot{e}/e$ as a function of the initial orbital
    eccentricity $e$. Each panel summarizes simulations for a different planetary mass $M_{\mathrm{p}}$, as represented by the labels. Points and coloured segmented lines
    distinguish different values of the accretion luminosity $L$ as specified in the legend.
    Using corresponding colours, the vertical arrows mark the equilibrium 
    eccentricity $e_{\mathrm{eq}}$ that the planets would asymptotically reach if they were
    evolving from a circular orbit at the given value of $L$. The horizontal zero line
    separates the regimes of eccentricity driving, $\dot{e}/e>0$, and damping, $\dot{e}/e<0$.
    Clearly, non-zero $e_{\mathrm{eq}}$ is reached in a substantial part of our parameter space.
    The case for $M_{\mathrm{p}}=0.5\,M_{\earth}$ is given in Fig.~\ref{fig:append_mp0.5}.
    We note that planets start migrating from $a_{1}=4.84\,\mathrm{au}$, i.e. from the minimum of $\eta$ (Fig.~\ref{fig:profiles}).
    }
    \label{fig:dedt}
\end{figure*}

\begin{figure}
    \centering
    \includegraphics{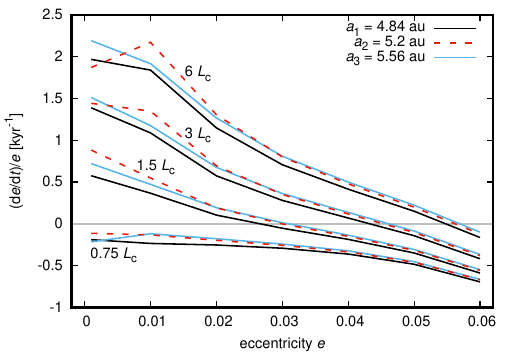}
    \caption{Eccentricity evolution rate $\dot{e}/e$ as a function of $e$
    for $M_{\mathrm{p}}=1\,M_{\earth}$ migrating
    from different initial semi-major axes:
    $a_{1}=4.84$ (black solid curve), $a_{2}=5.2$ (dashed red curve), and $a_{3}=5.56\,\mathrm{au}$ (blue solid curve). The labels mark individual sets of curves with 
    the respective value of $L$. For a given $L$, the curves are rather similar
    and therefore, $e_{\mathrm{eq}}$ is nearly independent of the initial position of the
    planet relative to the pressure bump.}
    \label{fig:dedt_acmpr}
\end{figure}

\begin{figure*}
    \centering
    \begin{tabular}{cc}
    \includegraphics[width=0.95\columnwidth]{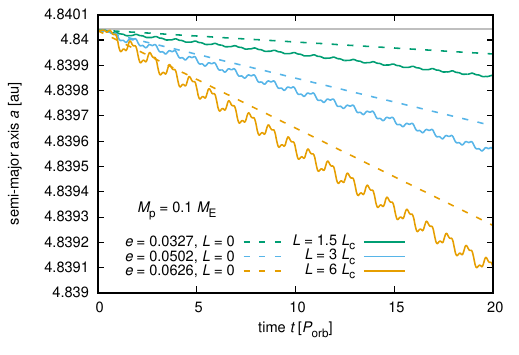} &
    \includegraphics[width=0.95\columnwidth]{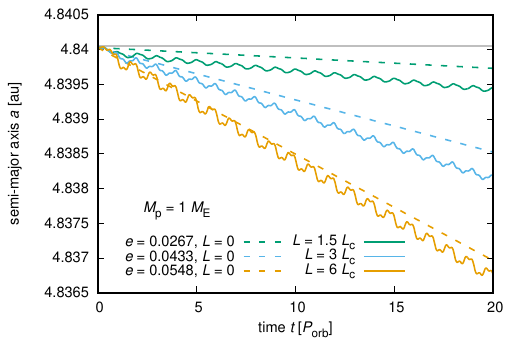} \\
    \includegraphics[width=0.95\columnwidth]{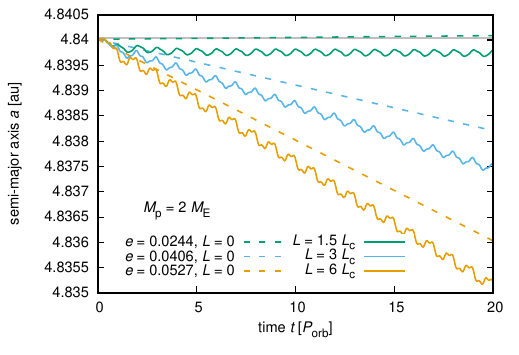} & 
    \includegraphics[width=0.95\columnwidth]{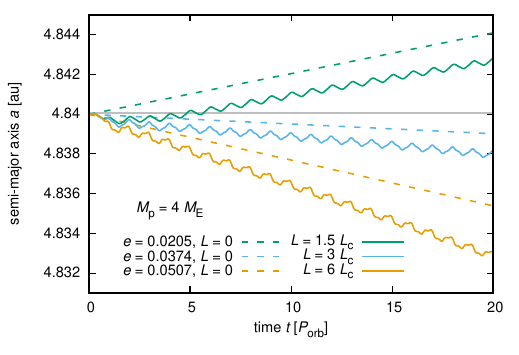} \\
    \end{tabular}
    \caption{Temporal evolution of the semi-major axis $a$ for planets of different
    masses (see the labels of individual panels) in the headwind-dominated regime of thermal torques. Solid curves follow the migration of
    luminous planets starting at non-zero eccentricities $e$ for which $\dot{e}=0$ (Fig.~\ref{fig:dedt}). The values of $e$ and $L$ are given in the plot legend.
    We remind the reader that the values of $e$ represent $e_{\mathrm{eq}}$ for all luminous cases with the exceptions of $M_{\mathrm{p}}=4\,M_{\earth}$ and $M_{\mathrm{p}}=2\,M_{\earth}$, $L=3\,L_{\mathrm{c}}$.
    Dashed curves are the expected migration tracks for non-luminous planets
    at the same $e$ (they were obtained by analyzing the power of disc-driven forces affecting
    a planet in a fixed eccentric orbit).
    The horizontal gray line represents the initial semi-major axis.
    For both luminous and non-luminous planets, the change in the migration rate (slope)
    is mostly driven by the change in the eccentricity, not by a modification of thermal torques themselves. The reader is referred to Fig.~\ref{fig:append_mp0.5} for $M_{\mathrm{p}}=0.5\,M_{\earth}$.
    }
    \label{fig:at_eeq}
\end{figure*}

\begin{figure*}
    \centering
    \begin{tabular}{cc}
    \includegraphics[width=0.95\columnwidth]{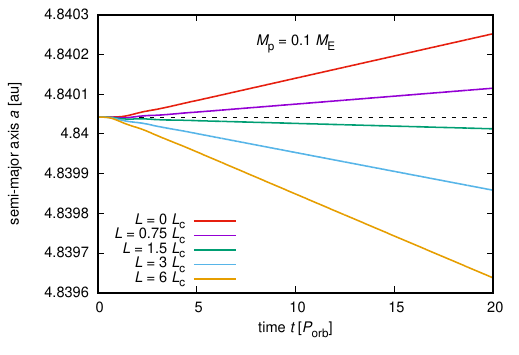} &
    \includegraphics[width=0.95\columnwidth]{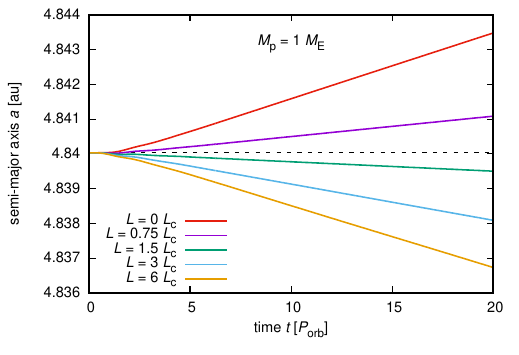} \\
    \includegraphics[width=0.95\columnwidth]{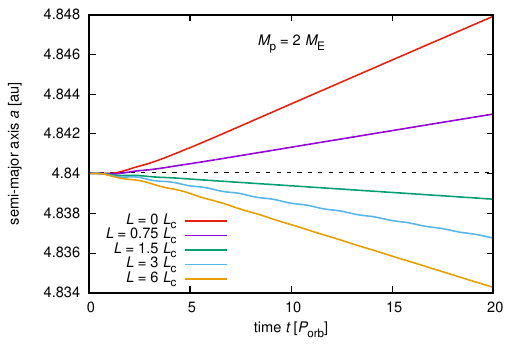} & 
    \includegraphics[width=0.95\columnwidth]{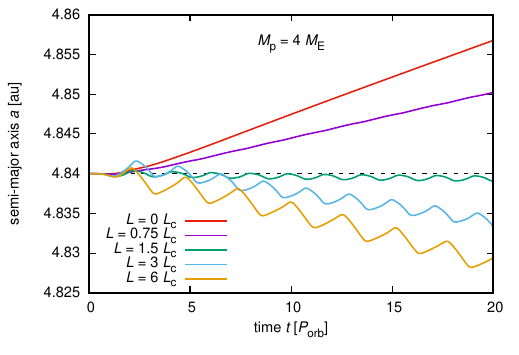} \\
    \end{tabular}
    \caption{Evolution of the semi-major axis $a$ for planets starting at $a_{1}$ and $e=0$
    (circular planets orbiting near the minimum of $\eta$).
    Individual panels are labelled with the planetary masses $M_{\mathrm{p}}$; the curves
    are coloured according to the luminosity of the planet $L$ (see the plot legend). 
    Unlike in typical sub-Keplerian discs, here the shear-dominated regime
    of thermal torques facilitates inward migration with increasing $L$.
    The reader is referred to Fig.~\ref{fig:append_mp0.5} for $M_{\mathrm{p}}=0.5\,M_{\earth}$.
    }
    \label{fig:at_e0}
\end{figure*}

\begin{figure}
    \centering
    \includegraphics[width=0.95\columnwidth]{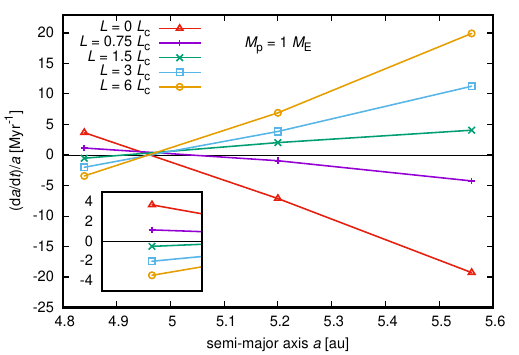}
    \caption{Evolution rate of the semi-major axis as a function of the 
    position with respect to the pressure bump for $M_{\mathrm{p}}=1\,M_{\earth}$.
    The planet starts at $e=0$ and thermal torques operate in the shear-dominated regime.
    Different points and colours represent different luminosities of the planet,
    as specified in the plot legend. The inset shows an enlarged detail of measurements
    at $a_{1}$. We point out that the influence of thermal
    torques on the resulting migration rate is reversed at $a_{1}=4.84\,\mathrm{au}$ compared to $a_{2}=5.2$ and $a_{3}=5.56\,\mathrm{au}$ due to the variations of the background pressure support.}
    \label{fig:dadt_e0_1me}
\end{figure}

\begin{figure*}
    \centering
    \begin{tabular}{cc}
    \includegraphics[width=0.95\columnwidth]{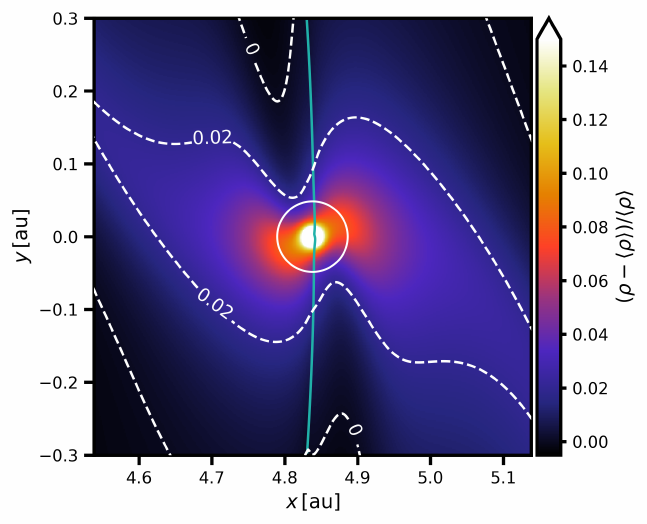} &
    \includegraphics[width=0.95\columnwidth]{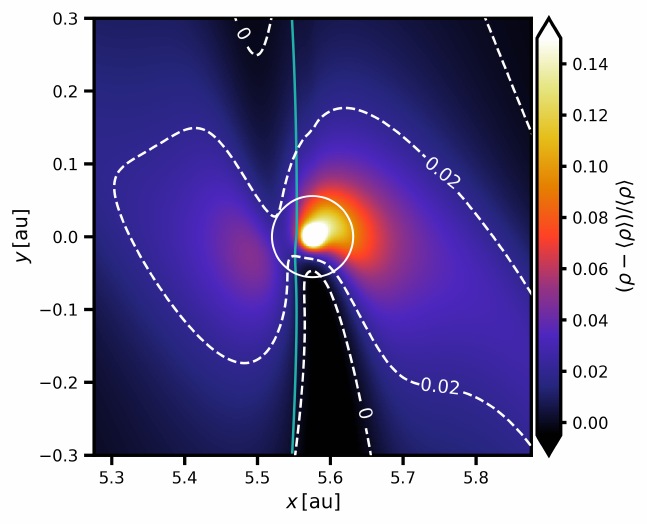} \\
    \end{tabular}
    \caption{Perturbation of the gas density $\rho$ relative to the azimuthally-averaged 
    gas density $\langle\rho\rangle$ in the disc midplane. The planet with $M_{\mathrm{p}}=1\,M_{\earth}$ is positioned in the centre of each panel and in the inertial frame, it would orbit in the $y>0$ direction. \emph{Left:} Situation at $a_{1}$ 
    where the disc is super-Keplerian. \emph{Right:} Situation at $a_{3}$ where the 
    disc is strongly sub-Keplerian.
    The white circle marks the Hill sphere, the light green curve marks the position where the disc material corotates
    with the planet ($v_{\theta}=0$ with respect to the planet), and white dashed curves are selected isocontours.
    The closeness of a given isocontour to the Hill sphere can be used as a measure
    of the asymmetry between the two underdense lobes (isocontours near a deeper lobe are prolated closer to the planet). The inner leading lobe dominates at $a_{1}$ while the outer trailing lobe dominates at $a_{3}$. The thermal torque is negative at $a_{1}$
    while it is positive at $a_{3}$.}
    \label{fig:gasdens}
\end{figure*}

\begin{table}
    \centering
    \begin{tabular}{ccc}
    \hline\hline
         Parameter & Notation/Scaling & Values \\
         \hline
         Planet mass & $M_{\mathrm{p}}/M_{\earth}$ & $0.1, 0.5, 1, 2, 4$\\
         Semi-major axis & $a/\mathrm{au}$ & $4.84, 5.2, 5.56$ \\
         Luminosity & $L/L_{\mathrm{c}}$ & $0, 0.75, 1.5, 3, 6$ \\
         Eccentricity & $e$ & $(0,0.1, 1, 2, 3, 4, 5, 6)\times 10^{-2}$ \\
         \hline
    \end{tabular}
    \caption{Parameter space explored in our simulations. The respective combinations
    of parameters are specified in individual sections. We remind the reader
    that $L_{\mathrm{c}}$ itself is a function of $M_{\mathrm{p}}$.}
    \label{tab:paramspace}
\end{table}

\subsection{Eccentricity excitation of luminous planets in the super-Keplerian bump region}
\label{sec:dedt}

Thermal forces can in principle change both the semi-major axis $a$
and eccentricity $e$ of migrating planets\footnote{Thermal forces can
also pump orbital inclinations \citep{Eklund_Masset_2017MNRAS.469..206E}, although
this effect is often quenched by the eccentricity growth and thus
not considered in our study.}. Therefore, to evaluate the migration of planets near
pressure bumps, it is necessary to establish whether $e$ can become non-zero and if so,
what is the equilibrium value $e\rightarrow e_{\mathrm{eq}}$ for which the
eccentricity driving and damping are balanced.

We start our investigation at $a_{1}=4.84\,\mathrm{au}$ where the 
region of outward (or stalled) migration is expected to exist and planets should become trapped
close to it (Fig.~\ref{fig:tqmap}). We explore the parameter space of planetary masses $M_{\mathrm{p}}/M_{\earth}=(0.1,0.5,1,2,4)$,
initial eccentricities $e(t=0)=(0.1,1,2,3,4,5,6)\times10^{-2}$, and luminosities $L/L_{\mathrm{c}}=(0.75, 1.5, 3, 6)$ that are scaled using the
critical luminosity \citep{Masset_2017MNRAS.472.4204M}
\begin{equation}
    L_{\mathrm{c}} = \frac{4\pi GM_{\mathrm{p}}\chi\rho}{\gamma} \, .
    \label{eq:L_crit}
\end{equation}
The critical luminosity is expected to separate the regimes of eccentricity damping and driving
for $L<L_{\mathrm{c}}$
and $L>L_{\mathrm{c}}$, respectively
\citep{Masset_VelascoRomero_2017MNRAS.465.3175M,Fromentau_Masset_2019MNRAS.485.5035F,Velasco-Romero_etal_2022MNRAS.509.5622V,Cornejo_etal_2023MNRAS.tmp..658C}.
By plugging equation (\ref{eq:L_crit}) into (\ref{eq:L}),
one finds that the mass doubling time necessary to reach $L_{\mathrm{c}}$
scales as $\tau\propto M_{\mathrm{p}}^{2/3}$.
For our span of planetary masses $M_{\mathrm{p}}/M_{\earth}=(0.1,0.5,1,2,4)$,
the luminosity becomes equal to $L_{\mathrm{c}}$ for $\tau\simeq(36,105,166,264,419)\,\mathrm{kyr}$,
respectively.
These values of $\tau$
are substantially larger with respect to our typical integration times, which cover several planetary orbits. It is therefore appropriate to keep the planetary masses fixed in our simulations and consider the luminosity as a free parameter.
Since our high-resolution radiative simulations are numerically demanding and the number
of parameters is relatively large, we start by performing simulations over $5$ orbital
periods of the planet (with a few exceptions specified below). Planets are smoothly introduced in the simulations by ramping $M_{\mathrm{p}}$
from zero to its parametric value over the first orbital time-scale.

Fig.~\ref{fig:dedt} shows the eccentricity evolution rate $\dot{e}/e$ as a function
of the initial orbital eccentricity for all our parameters (see Fig.~\ref{fig:append_mp0.5}
for $M_{\mathrm{p}}=0.5\,M_{\earth}$). The time derivative of the eccentricity $\dot{e}$
was obtained by fitting a linear function to the time series of $e(t)$ that were
recorded during our simulations. The fit was performed
over the last three orbits and $\dot{e}$ was determined from its slope.
For several cases with the lowest eccentricity $e=10^{-3}$, the behaviour
of $e(t)$ was deviating from a linear trend on the time-scale of three orbits.
We thus extended these specific cases to twenty orbital periods
and we performed the linear fit over the last eighteen orbits\footnote{The linear fitting
technique adequately describes $\dot{e}/e$
as long as the planet mass
and luminosity remain constant.
Nevertheless, when fluctuations 
in the shape of thermal lobes are present (see Section~\ref{sec:fluct}),
the time interval of the fit has to be larger
than the time-scale of the fluctuations.
The presence of such fluctuations
is the main reason why some of our 
simulations at $e=10^{-3}$ had to be
prolonged.}.

Similarly to \cite{Eklund_Masset_2017MNRAS.469..206E} or \cite{Velasco-Romero_etal_2022MNRAS.509.5622V},
Fig.~\ref{fig:dedt} allows us to find the equilibrium asymptotic eccentricity $e_{\mathrm{eq}}$
that the planets would reach. If $\dot{e}<0$ for $e=10^{-3}$, we take $e_{\mathrm{eq}}=0$.
Such cases represent planets that, when starting on circular orbits, feel the eccentricity
damping straightaway, which prevents them from any eccentricity growth.
If, on the other hand, $\dot{e}>0$ for $e=10^{-3}$, we find $e_{\mathrm{eq}}$ by applying
linear interpolation to our measurements and identifying the point where $\dot{e}=0$.
In these cases, planets starting at $e<e_{\mathrm{eq}}$ experience eccentricity driving
while planets starting at $e>e_{\mathrm{eq}}$ experience eccentricity damping.
Equilibrium eccentricities $e_{\mathrm{eq}}$ are marked by arrows in Fig.~\ref{fig:dedt}
and summarized in Table~\ref{tab:e_eq}.

We see that the planets remain circular as long as their luminosity is sub-critical.
Once the luminosity becomes super-critical, $e_{\mathrm{eq}}$ reaches values
of the order of $h=H/r$ \citep{Chrenko_etal_2017A&A...606A.114C} in most cases.
As for other trends, larger $L$ implies larger $e_{\mathrm{eq}}$ for fixed $M_{\mathrm{p}}$,
while smaller $M_{\mathrm{p}}$ implies larger $e_{\mathrm{eq}}$ for fixed $L$,
in good agreement with \cite{Velasco-Romero_etal_2022MNRAS.509.5622V}.
In the case with $M_{\mathrm{p}}=2\,M_{\earth}$ and $L=3\,L_{\mathrm{c}}$, as well
as for all cases with $M_{\mathrm{p}}=4\,M_{\earth}$, we find $e_{\mathrm{eq}}=0$
even for the super-critical $L$.
We attribute this behaviour to the non-linearities that arise for increasing
planetary masses for which the advective redistribution
of hot gas close to the planet does not reach a steady state
\citep{Chrenko_Lambrechts_2019}.
These cases are covered in greater detail in Section~\ref{sec:fluct}.
For $M_{\mathrm{p}}\leq1\,M_{\earth}$, however, we conclude that $L_{\mathrm{c}}$
\citep[that was originally derived from the linear perturbation theory in discs with thermal diffusion;][]{Masset_2017MNRAS.472.4204M}
is a robust separation for the eccentricity damping and excitation
even in 3D radiative discs.

\begin{table}
    \centering
    \begin{tabular}{cccccc}
    \hline\hline
    $L$ & \multicolumn{5}{c}{$e_{\mathrm{eq}}$ for $M_{\mathrm{p}}/M_{\earth}=$} \\
                       & 0.1 & 0.5 & 1 & 2 & 4 \\ 
    \hline
    0.75 $L_{\mathrm{c}}$ & 0 & 0 & 0 & 0 & 0 \\
    1.5 $L_{\mathrm{c}}$ & 0.033 & 0.029 & 0.027 & 0.024 & 0 \\
    3 $L_{\mathrm{c}}$ & 0.05 & 0.046 & 0.043 & 0 & 0 \\
    6 $L_{\mathrm{c}}$ & 0.063 & 0.057 & 0.055 & 0.053 & 0 \\
    \hline
    \end{tabular}
    \caption{Equilibrium eccentricities $e_{\mathrm{eq}}$ (\emph{columns 2--6})   
    for various planetary masses $M_{\mathrm{p}}$ and luminosities $L$ (\emph{column 1}).
    For $L>L_{\mathrm{c}}$, planets acquire orbits with
    substantial eccentricities in the majority of our parameter space.
    }
    \label{tab:e_eq}
\end{table}

\subsection{Headwind-dominated regime}
\label{sec:hw}

Section~\ref{sec:dedt} reveals that
$e_{\mathrm{eq}}\sim h$ for the majority of our parameter space.
These relatively large eccentricities place the planets firmly to the headwind-dominated
regime of thermal torques \citep{Eklund_Masset_2017MNRAS.469..206E} 
for which the two-lobed density perturbation \citep{Benitez-Llambay_etal_2015Natur.520...63B}
becomes replaced by a single hot trail \citep{Chrenko_etal_2017A&A...606A.114C}.
Here, we focus on the headwind-dominated regime in greater detail.

As a first step, we need to establish whether the magnitude of eccentricity excitation
from Fig.~\ref{fig:dedt} is universal or rather dependent on the local conditions
within the pressure bump. To this point, we repeated the measurement from Fig.~\ref{fig:dedt}
for $M_{\mathrm{p}}=1\,M_{\earth}$ but this time we placed the planet at
initial semi-major axes $a_{2}=5.2$ and $a_{3}=5.56\,\mathrm{au}$ to
investigate $\dot{e}$ at the bump centre and at the maximum of $\eta$, respectively.

Fig.~\ref{fig:dedt_acmpr} compares the eccentricity evolution rate at different
locations in the pressure bump. Although we detect marginal differences at
very low $e$, the values obtained at fixed $L$ are rather similar, which makes them independent 
of $a$. Therefore, the ability of a planet to reach the headwind-dominated
regime with non-zero $e_{\mathrm{eq}}$ does not depend on its position with respect to
the pressure bump. Once the planet develops non-zero $e_{\mathrm{eq}}$ anywhere, it will
tend to maintain it despite its radial migration.

Next, we explore the migration in the headwind-dominated regime. Focusing solely
on $a_{1}$ again, we let the planets migrate using $e_{\mathrm{eq}}$ from Table~\ref{tab:e_eq}
as their initial eccentricities.
For the cases with $L\geq1.5\,L_{\mathrm{c}}$ and $M_{\mathrm{p}}\geq 2\,M_{\earth}$ that exhibit $e_{\mathrm{eq}}=0$ in Table~\ref{tab:e_eq}, we start from their largest $e$ that
would lead to $\dot{e}=0$ in Fig.~\ref{fig:dedt}. The simulation time-scales for migrating
planets are equal to 20 orbital periods.

Fig.~\ref{fig:at_eeq} (solid curves) shows the resulting temporal evolution of semi-major axes. Clearly,
planets with super-critical luminosities often abandon the outward migration predicted
by Fig.~\ref{fig:tqmap} and they predominantly switch to inward migration. We only detect stalled
migration for $M_{\mathrm{p}}=2\,M_{\earth}$ and $L=1.5\,L_{\mathrm{c}}$, outward
migration for $M_{\mathrm{p}}=4\,M_{\earth}$ and $L=1.5\,L_{\mathrm{c}}$, and we point
out that the migration of $M_{\mathrm{p}}=0.1\,M_{\earth}$ is generally slow because of its very
low mass (see the extent of the vertical axis in Fig.~\ref{fig:at_eeq}). 

Since the inward migration becomes faster with increasing $L$, is is natural to ask whether
this effect is somehow regulated by the thermal torques themselves or whether it is 
driven by the Lindblad and corotation torques operating at increased eccentricities.
To find the answer, we performed simulations with non-luminous planets that are subject
to cold thermal torques in radiative discs
\citep{Lega_etal_2014MNRAS.440..683L,Masset_2017MNRAS.472.4204M}. We placed
these planets on fixed eccentric orbits with the same span of $e$ as used for their luminous
counterparts. The purpose of fixing the orbits is to prevent the eccentricity damping.
To predict the migration outcome, we measured the torque $\Gamma$ and power $P$ exerted by
the gravitational disc forces, then we averaged these values over the last ten orbital periods,
and we estimated \citep[e.g.][]{Bitsch_Kley_2010A&A...523A..30B}
\begin{equation}
\frac{\dot{a}}{a} = \frac{2a}{GM_{\star}M_{\mathrm{p}}}P \, ,
\label{eq:dadt}
\end{equation}
which expresses the evolution rate of the semi-major axis from the change in the orbital energy.

By integrating Equation~(\ref{eq:dadt}), we obtained the dashed curves shown in Fig.~\ref{fig:at_eeq}.
Focusing on the change in the slope of $a(t)$ curves,
it becomes clear that the cases with $L=0$ exhibit the same change of slope
with the increase of $e$ as the cases with $L\neq 0 $. Although there is a small systematic offset between the $L=0$ and $L\neq 0 $ curves, it does not seem to strongly depend on the
actual value of $L$ (in contrast to Fig.~\ref{fig:at_e0} discussed later).
Therefore, we obtain an important result: The actual migration rate for $e_{\mathrm{eq}}$
corresponding to the headwind-dominated regime is not controlled by thermal torques
themselves, these are only responsible for keeping $e$ pumped up. 
Instead, the migration rate is controlled by the standard Lindblad and corotation torques.
Since the latter experiences exponential quenching with increasing $e$ \citep{Fendyke_Nelson_2014MNRAS.437...96F}, the inward
migration in Fig.~\ref{fig:at_eeq} becomes faster for larger $e$ as the Lindblad torque
becomes more and more dominant.

The implications of this section are the following. The excitation of $e_{\mathrm{eq}}\sim h$
is a global effect independent of planet position in the disc and at this level
of eccentricity, the migration is governed mostly by the Lindblad and corotation torques.
Then, if a luminous planet is found to exhibit inward migration at $a_{1}$ where Fig.~\ref{fig:tqmap}
predicts outward (or stalled) migration and where the corotation torque is the strongest,
it will be subject to inward migration in our entire simulated disc because the corotation
torque can only become weaker away from $a_{1}$ and the Lindblad torque is always negative.
Hence, for the inward-migrating planets from Fig.~\ref{fig:at_eeq}, the migration trap 
at the pressure bump does not exist.

\subsection{Shear-dominated regime}
\label{sec:shear}

Let us now turn out attention to the migration at $e=0$ for which the
thermal torques operate in the shear-dominated regime.
This regime becomes important when $L<L_{\mathrm{c}}$ (Table~\ref{tab:e_eq}),
or when the behaviour of thermal torques becomes highly non-linear
(e.g. for our case $M_{\mathrm{p}}=4\,M_{\earth}$), and we also expect that 
planets can temporarily evolve in this regime if they are born in circular orbit
before having their $e$ excited.

We performed simulations of both cold and luminous planets ($L/L_{\mathrm{c}}=(0,0.75,1.5,3,6)$) evolving from $e=0$ and starting at three distinct positions $a_{1}$,
$a_{2}$, and $a_{3}$ spread across the pressure bump. The simulations covered 20 orbital
periods again. Fig.~\ref{fig:at_e0} depicts the migration tracks at $a_{1}$. The migration is outward,
and thus in accordance with the migration map in Fig.~\ref{fig:tqmap}, as long as $L<L_{\mathrm{c}}$ and we also find one case of stalled migration for $M_{\mathrm{p}}=4\,M_{\earth}$, $L=1.5\,L_{\mathrm{c}}$. 
For the remaining cases, however, the migration becomes directed inwards and surprisingly,
its rate becomes faster with increasing $L$.
This is in striking contrast to the usual behaviour of thermal torques in
simple power-law discs \citep{Benitez-Llambay_etal_2015Natur.520...63B}
where increasing $L$ typically makes the migration tracks more and more outward.

To further highlight this finding, Fig.~\ref{fig:dadt_e0_1me} shows the evolution
rate of the semi-major axis $\dot{a}/a$ for $M_{\mathrm{p}}=1\,M_{\earth}$ 
at all three initial semi-major axes $a_{1}$, $a_{2}$, and $a_{3}$.
The behaviour at $a_{1}$ is as described in the previous paragraph. 
The behaviour at $a_{2}$ and $a_{3}$ is antisymmetric, in a sense that
the fastest inward migration (i.e. the most negative total torque) is found for $L=0$
and then it slows down and reverses as $L$ increases (i.e. the torque receives gradual
positive boosts each time $L$ grows).

We infer that these trends are related to the local disc rotation and pressure support.
The locations $a_{2}$ and $a_{3}$ have $\eta>0$ (Fig.~\ref{fig:profiles})
and the local rotation is sub-Keplerian.
Therefore, the exact corotation between the disc material and the planet is offset inwards
with respect to the planet and the resulting two-lobed underdensity is asymmetric because
the disc shear is more efficient in advecting the hot gas into the outer lobe, trailing
the orbital motion of the planet.
Since $\eta$ is larger at $a_{3}$, so is the offset of the planet from the corotation 
and the thermal torques are more prominent compared to $a_{2}$. This is in accordance with
\cite{Benitez-Llambay_etal_2015Natur.520...63B}.

However, $\eta<0$ at $a_{1}$ (super-Keplerian rotation)
and the same reasoning implies that the exact corotation is
located outwards from the planet and the inner underdense lobe that leads the orbital motion of the planet will dominate over the outer one. In other words, the lack of gas ahead of the planet means that the planet will be robed of angular momentum by the gas trailing it. The total torque will therefore become more negative, which facilitates the shift towards inward migration found in Figs.~\ref{fig:at_e0} and \ref{fig:dadt_e0_1me} at the minimum of $\eta$.

As a verification of our claims, Fig.~\ref{fig:gasdens} shows the density perturbation 
in the disc midplane near an Earth-mass planet. The density perturbation $(\rho-\langle\rho\rangle)/\langle\rho\rangle$ is computed with respect to the azimuthal average $\langle\rho\rangle$.
By tracing the isocontours, one can see that the excavation of the inner lobe 
is larger at $a_{1}$ (left panel in Fig.~\ref{fig:gasdens}), while the outer lobe dominates
at $a_{3}$ (right panel in Fig.~\ref{fig:gasdens}). This asymmetry depends on where
the disc material corotates with the planet, as indicated by the light green curve in Fig.~\ref{fig:gasdens}.
We point out that the dependence of thermal torques on the corotation offset has already been described in previous works \citep{Benitez-Llambay_etal_2015Natur.520...63B,Masset_2017MNRAS.472.4204M,Chametla_Masset_2021MNRAS.501...24C}, however,
the case when the corotation is located farther out with respect to the planet
was not considered in detail or it was explicitly regarded as unrealistic \citep{Benitez-Llambay_etal_2015Natur.520...63B}.
Here, instead, we demonstrate that the impact of such an outward corotation offset, which
inherently occurs near pressure bumps (left panel in Fig.~\ref{fig:gasdens}), can be substantial.

Before summarizing this section, let us also point out that while the evolution of the semi-major axes can be reverted depending on $\eta$,
the eccentricity evolution rate is affected only weakly as we saw in Fig.~\ref{fig:dedt_acmpr}.
An extended analysis of this fact for the shear-dominated regime is provided in
Appendix~\ref{sec:app_egrowth}.

The implications of this section are best inferred from Fig.~\ref{fig:dadt_e0_1me},
which reveals a radius of convergent migration (roughly at $4.95\,\mathrm{au}$)
for $L<L_{\mathrm{c}}$ but 
leads to divergent migration for $L>L_{\mathrm{c}}$.
Therefore, if planets with super-critical luminosities remain circular (which happens
only for a limited subset of our parameter space), they are again expected 
to migrate away from the pressure bump. If these planets were to start at $a_{1}$,
they could possible get trapped near $a\simeq4.67\,\mathrm{au}$ where $\eta$ becomes positive
again and thus we can expect another change in the sign of $\dot{a}$ (not explored in our Fig.~\ref{fig:dadt_e0_1me}).
If these planets were to start at $a_{2}$ or $a_{3}$, they would migrate outwards.

\begin{figure*}
    \centering
    \begin{tabular}{cc}
    \includegraphics[width=0.95\columnwidth]{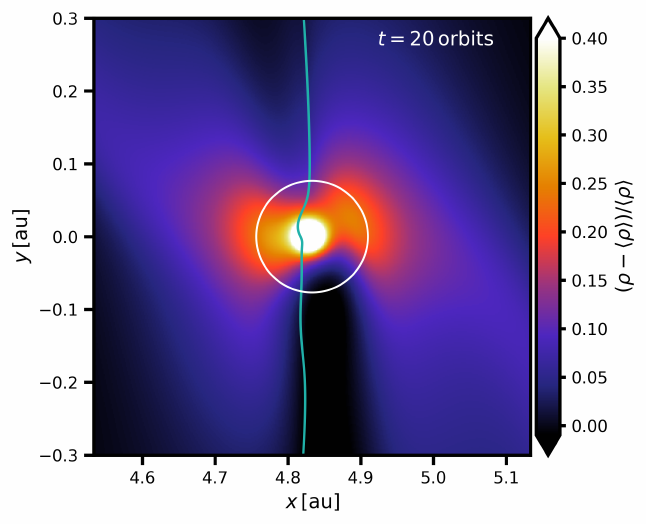} &
    \includegraphics[width=0.95\columnwidth]{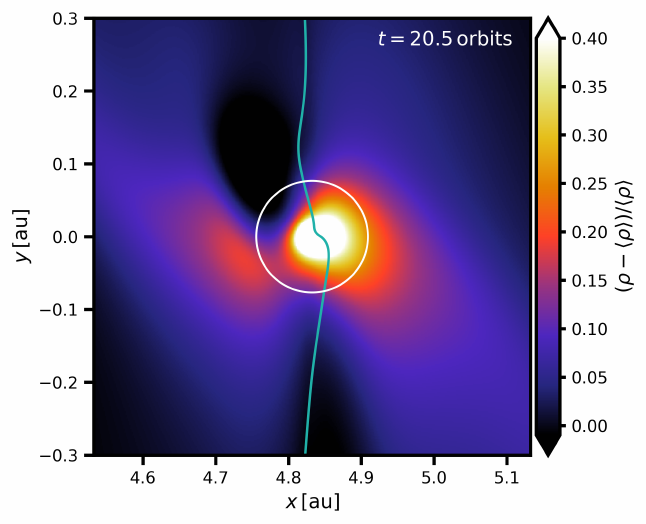} \\
    \includegraphics[width=0.95\columnwidth]{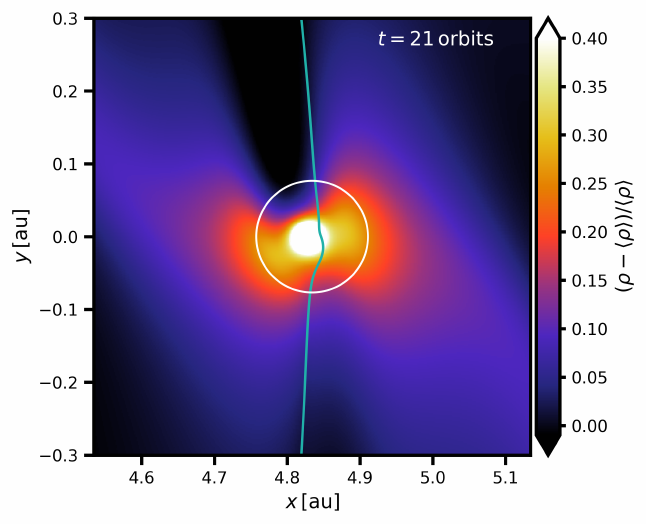} & 
    \includegraphics[width=0.95\columnwidth]{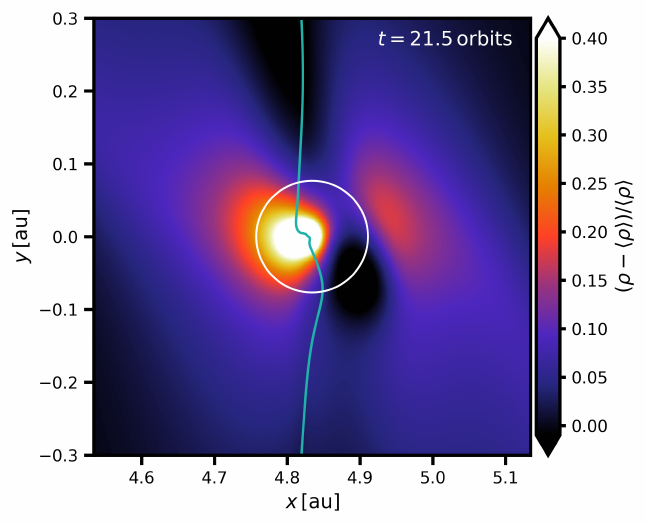} \\
    \end{tabular}
    \caption{As in Fig.~\ref{fig:gasdens} but for $M_{\mathrm{p}}=4\,M_{\earth}$
    at $a_{1}$ with $L=6\,L_{\mathrm{c}}$. The two \emph{rows} cover two consecutive orbital cycles
    (individual \emph{panels} are labelled with corresponding simulation times $t$).
    The fluctuations suggest that the circumplanetary flow
    never reaches a steady state, similarly to \protect\cite{Chrenko_Lambrechts_2019}.}
    \label{fig:gasdens_osc}
\end{figure*}

\subsection{Fluctuating thermal disturbance}
\label{sec:fluct}

When assessing the equilibrium eccentricities, we saw in Fig.~\ref{fig:dedt} that 
while the planets with $M_{\mathrm{p}}\leq 1\,M_{\earth}$ exhibit an orderly and mildly decreasing dependence
of $\dot{e}$ on $e$, our most massive planets exhibit a break near the smallest eccentricities.
Consequently, these planets are capable of remaining circular even at super-critical luminosities (Table~\ref{tab:e_eq}). Similarly, when focusing on the migration of these planets
in the circular case, we saw that the semi-major axis evolution is accompanied by short-term oscillations
that are best apparent for $M_{\mathrm{p}}=4\,M_{\earth}$ and $L\gtrsim1.5\,L_{\mathrm{c}}$ (Fig.~\ref{fig:at_e0}, bottom right).

By investigating the gas evolution for these peculiar cases, we found that they exhibit fluctuations
of the thermal disturbance. These fluctuations were first reported in \cite{Chrenko_Lambrechts_2019}
and they are driven by a complex reconfiguration of the 3D circumplanetary flow.
An interesting fact is that while \cite{Chrenko_Lambrechts_2019} found the presence of these fluctuations in a setup with temperature-dependent disc opacities, here we obtain the same behaviour
in a constant-opacity disc.

Fig.~\ref{fig:gasdens_osc} shows the temporal evolution of the gas density perturbation
for $M_{\mathrm{p}}=4\,M_{\earth}$ and $L=6\,L_{\mathrm{c}}$ during two orbits of the planet.
At $t=20$ orbits (top left panel), the outer rear lobe dominates.
After one half of the orbit (top right panel), the outer rear lobe shrinks while the inner front lobe becomes more pronounced.
At $t=21$ orbits (bottom left panel), the inner front lobe dominates. The cycle returns to the beginning
shortly after $t=21.5$ orbits (bottom right panel).

The realism of the fluctuating thermal disturbance is still somewhat unclear because \cite{Chrenko_Lambrechts_2019} showed
that its occurrence is favoured in discs with vertically steep temperature gradients.
Realistic protoplanetary discs that are stellar-irradiated have shallower vertical temperature
gradients than discs heated purely by viscous friction (considered here), future work should 
therefore examine the occurrence of fluctuating thermal disturbances in stellar-irradiated discs as well.
However, the fact that the fluctuations
naturally appear in our constant-opacity setup with much better resolution than \cite{Chrenko_Lambrechts_2019}
provides additional independent indication of their robustness.

\begin{figure}
    \centering
    \includegraphics[width=0.95\columnwidth]{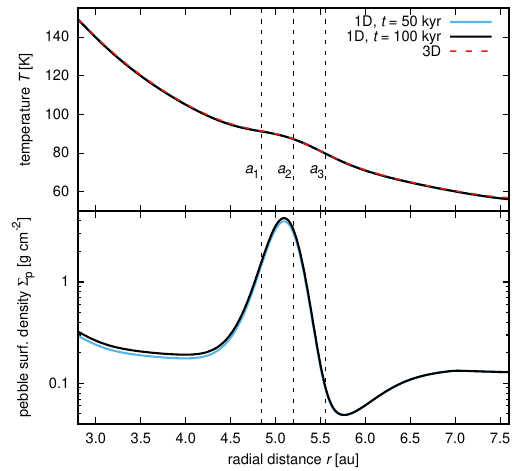}
    \caption{Radial profile of the gas temperature $T$ (\emph{top})
    and pebble surface density $\Sigma_{\mathrm{p}}$ (\emph{bottom})
    during the evolution (blue and black curves) of our bi-fluid 1D model.
    Red dashed curve in the \emph{top panel} shows the equilibrium
    profile of our 3D disc (see Fig.~\ref{fig:profiles}) for reference.
    The vertical dashed lines mark the positions $a_{1}$, $a_{2}$,
    and $a_{3}$ used in our simulations with planets.}
    \label{fig:1d_profiles}
\end{figure}

\begin{figure}
    \centering
    \includegraphics[width=0.95\columnwidth]{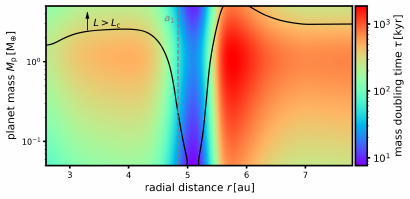}
    \caption{Mass doubling time $\tau$ due to pebble accretion
    as a function of the planet mass $M_{\mathrm{p}}$ and 
    radial distance $r$ in the disc. The region above the solid
    black curve is where the luminosity of accreting planets
    becomes super-critical, $L>L_{\mathrm{c}}$. The vertical
    dashed line marks the location $a_{1}$. Planets
    are accreting pebbles with $\mathrm{St}=0.07$ and the pebble flux
    at the outer boundary is $10^{-4}\,M_{\earth}\,\mathrm{yr}^{-1}$.}
    \label{fig:doubl_time_map}
\end{figure}

\subsection{Luminosity reached by pebble accretion}
\label{sec:peb_acc}

In previous sections, the luminosity $L$ of accreting planets
was a free parameter unrelated to any physical process.
The aim of this section is to determine the values of $L$
that can be reached by pebble accretion \citep{Ormel_Klahr_2010A&A...520A..43O,Lambrechts_Johansen_2012A&A...544A..32L},
which is thought to be a major
accretion channel in many planet formation scenarios \citep[e.g.][]{Lambrechts_etal_2014A&A...572A..35L,Bitsch_etal_2019A&A...624A.109B,Venturini_etal_2020A&A...644A.174V,Broz_etal_2021NatAs...5..898B}.

We constructed a simple model for a coupled evolution of gas and pebbles. We solved
2D vertically-integrated continuity and Navier-Stokes equations for a bi-fluid mixture of gas and
pebbles using Fargo3D again
and we additionally assumed axial symmetry, thus making the model effectively 1D.
Gas and pebbles were aerodynamically coupled \citep[e.g.][]{Chametla_Chrenko_2022MNRAS.512.2189C} and the turbulent pebble diffusion was taken into account
as well \citep{Weber_etal_2019ApJ...884..178W}.
To maintain the disc non-isothermal and thus comparable to our 3D model,
we included a one-temperature gas energy equation in the form of \cite{Chrenko_etal_2017A&A...606A.114C}
while neglecting the stellar irradiation term. Unlike in \cite{Chrenko_etal_2017A&A...606A.114C}
(see equation~11 of that paper), we did not include any opacity correction in the 
vertical optical depth \citep{Hubeny_1990ApJ...351..632H}. The disc parameters
were the same as in the case of our 3D disc.

The 1D simulation was started by an iterative hydrostatic relaxation similar to Section~\ref{sec:inicond}.
Once the gas temperature equilibrated, we added the pebble component by assuming a constant
pebble flux $\dot{M}_{\mathrm{F}}=10^{-4}\,M_{\earth}\,\mathrm{yr}^{-1}$ through the disc:
\begin{equation}
    \Sigma_{\mathrm{p}} = - \frac{\dot{M}_{\mathrm{F}}}{2\pi r v_{r,\mathrm{p}}} \simeq \frac{\dot{M}_{\mathrm{F}}}{4\pi r^{2}\mathrm{St}\eta \Omega_{\mathrm{K}}} \, ,
    \label{eq:sigma_p}
\end{equation}
where $\Sigma_{\mathrm{p}}$ is the initial pebble surface density, $v_{r,\mathrm{p}}$ is the radial drift velocity of pebbles and $\mathrm{St}=0.07$ is the Stokes number (dimensionless stopping time) of pebbles \citep{Nakagawa_etal_1986Icar...67..375N}. For the sake of pebble initialization, we used
$\eta$ corresponding to a bump-free disc (because Equation~\ref{eq:sigma_p} would lead to $\Sigma_{\mathrm{p}}<0$ for $\eta<0$ inside the bump, which would be unphysical). The value of $\mathrm{St}$ was derived in a bump-free disc again,
assuming that the dominant physical radius of pebbles is drift-limited \citep{Lambrechts_Johansen_2014A&A...572A.107L,Chrenko_etal_2017A&A...606A.114C}.
We evolved the disc for $8,500$ orbital time-scales, which evaluates roughly to $t\simeq100\,\mathrm{kyr}$.
To maintain the pebble flux entering the disc uniform, we damped $\Sigma_{\mathrm{p}}$ to its initial
value close to the outer boundary.

The result of our simulation is given in Fig.~\ref{fig:1d_profiles}. First, the top panel demonstrates
that the temperature profile of the 1D disc is indeed
comparable to our more advanced 3D model.
Second, the bottom panel shows the radial profile of $\Sigma_{\mathrm{p}}$
and reveals that our parametrization of the pressure bump does not create a perfect barrier to the pebble flux.
This is because the region of $\eta<0$ is relatively narrow (see Fig.~\ref{fig:profiles})
and thus, once the mass loading of the bump by pebbles becomes sufficient, the pebbles start
to penetrate the bump by means of turbulent diffusion. Due to the latter, the radial flux
of pebbles inwards from the bump remains non-zero.
Since $\Sigma_{\mathrm{p}}$ is not evolving
substantially between $t=50$ and $100\,\mathrm{kyr}$, it is clear that an equilibrium state was reached.

Having the knowledge of $\Sigma_{\mathrm{p}}$, we calculated the pebble accretion rate for a range
of planetary masses across the disc. The calculation followed the recipe of \cite{Liu_Ormel_2018A&A...615A.138L} that was reformulated by \cite{Jiang_Ormel_2023MNRAS.518.3877J}
for the environment of pressure bumps.
We neglected any vertical stirring of pebbles (and thus also the 3D regime of pebble accretion)
and we assumed circular orbits of planets. The results are shown in Fig.~\ref{fig:doubl_time_map}
in terms of the mass doubling time $\tau$. Clearly, the accumulation of pebbles at the bump
boosts the accretion efficiency (shortens $\tau$) and the growth of planets can become rapid.
In the light of our study, however, we can see that growing planets easily exceed the critical
luminosity $L_{\mathrm{c}}$. At $a_{1}$, this happens for $M_{\mathrm{p}}\simeq 0.2\,M_{\earth}$.
Consequently, planets with $M_{\mathrm{p}}\gtrsim 0.2\,M_{\earth}$ would undergo eccentricity excitation,
they would enter the headwind-dominated regime, and they would start migrating inwards.

\section{Discussion}
\label{sec:discus}

\subsection{On the viability of model parameters}

The full parameter space of our model is extensive and we thus only focused on exploring several specific dependencies. We have already explained our rationale behind selecting parameters
for the pressure bump and planetary luminosity in Sections~\ref{sec:bump_param} and \ref{sec:dedt}, respectively. However, it is important to discuss the remaining model parameters, namely,
the opacity $\kappa=1\,\mathrm{cm}^{2}\,\mathrm{g}^{-1}$ and the kinematic viscosity $\nu=10^{15}\,\mathrm{cm}^{2}\,\mathrm{s}^{-1}$. These values were adopted from \cite{Eklund_Masset_2017MNRAS.469..206E} in order to establish a connection between our study
and prior research. Nevertheless, one should bear in mind that $\kappa$ and $\nu$ can vary over orders of magnitude in realistic protoplanetary discs.

For less opaque protoplanetary discs, the efficiency of thermal torques decreases, as numerically demonstrated by \cite{Benitez-Llambay_etal_2015Natur.520...63B} and analytically shown
by \cite{Masset_2017MNRAS.472.4204M}.
This torque reduction occurs
because a lower opacity leads to a larger thermal diffusivity (Equation~\ref{eq:chi}) 
while the hot thermal torque component scales as $\propto$$\chi^{-3/2}$ in the linear regime.
In the light of our study, however, we must also ask how the eccentricity excitation is affected.
By utilizing the results from linear theory,
we can roughly estimate that the eccentricity will grow as
long as (i) the response time of the thermal driving $t_{\mathrm{th}}$ is shorter than the time-scale of wave-induced damping
$t_{\mathrm{wave}}$ \citep{Tanaka_Ward_2004ApJ...602..388T} 
and (ii) the luminosity remains super-critical. Both of these conditions 
are dependent on $\chi$ and thus influenced by $\kappa$. Regarding (i), \cite{Fromentau_Masset_2019MNRAS.485.5035F} found
\begin{equation}
    \frac{t_{\mathrm{th}}}{t_{\mathrm{wave}}} = \sqrt{\frac{\pi}{2}}\frac{1}{\gamma\left(\gamma-1\right)}\frac{\lambda_{\mathrm{c}}}{H} \, ,
    \label{eq:t_th_vs_t_wave}
\end{equation}
and we recall that $\lambda_{\mathrm{c}}=0.1H$ in our model. To prevent the eccentricity growth of luminous planets born
in circular orbits, $\kappa$ would need to decrease sufficiently to expand $\lambda_{\mathrm{c}}\propto\sqrt{\chi}$ by an
order of magnitude. Nevertheless, determining the critical value of $\kappa$ at which the wave-induced damping dominates is not straightforward
because not only $\chi\propto\kappa^{-1}$, but also $\chi\propto T^{4}$ and we expect that the equilibrium temperature profile of a disc with lower opacity would be cooler. Regarding (ii), lower $\kappa$ (and consequently larger $\chi$) would elevate the critical luminosity threshold $L_{\mathrm{c}}$.
In other words, larger accretion rates (lower mass doubling times $\tau$) would be required to initiate the eccentricity growth.

To assess whether the value of $\kappa=1\,\mathrm{cm}^{2}\,\mathrm{g}^{-1}$ is appropriate
for the studied disc region, we compared it with the Rosseland opacities based on the DIANA standard for the dust composition \citep{Woitke_etal_2016A&A...586A.103W} within the temperature range of our disc. Appendix~\ref{sec:opa} and Fig.~\ref{fig:opa} show that these more detailed opacities are similar to $\kappa=1\,\mathrm{cm}^{2}\,\mathrm{g}^{-1}$ at the radial positions
$a_{1}$--$a_{3}$, which we have considered as the initial locations for the planets.
We admit, however, that the planets themselves are surrounded by temperature peaks when they
release the accretion heat (for instance, $M_{p}=1\,M_{\earth}$ at $L=3\,L_{\mathrm{c}}$ exhibits a temperature peak exceeding $200\,\mathrm{K}$)
and our constant-opacity model is certainly a simplification within these local temperature maxima.

Concerning the viscosity,
the value of $\nu=10^{15}\,\mathrm{cm^{2}}\,\mathrm{s}^{-1}$ translates to the Shakura-Sunyaev viscosity \citep{Shakura_Sunyaev_1973A&A....24..337S} of $\alpha\simeq3.5$--$4\times10^{-3}$ across the pressure bump, which is relatively large and requires a substantial level of turbulent
stress to operate. Such a stress might be difficult to achieve because a hydrodynamic turbulence
is typically weaker
\citep[e.g.][]{Nelson_etal_2013MNRAS.435.2610N,Klahr_etal_2018haex.bookE.138K,Pfeil_Klahr_2019ApJ...871..150P}
and a magneto-hydrodynamic turbulence in the midplane at several au is likely to be suppressed \citep[e.g.][]{Turner_etal_2014prpl.conf..411T,Lesur_etal_2022arXiv220309821L}.
At lower and vanishing viscosities, we think that the response of planets to the thermal
perturbation shown in our study remains valid
because \cite{Velasco-Romero_etal_2022MNRAS.509.5622V} reported similar
eccentricity evolution in models of inviscid power-law discs with thermal diffusion.
Additional effects, however, can be expected in discs with pressure bumps
because in the low-viscosity limit, embedded planets might perturb the bumps
more prominently and they might even destabilize them. This problem is left for future work.

\subsection{Implications for planet formation at pressure bumps}

Pressure bumps have been proposed as efficient sites of planet accretion 
in several recent works (Section~\ref{sec:intro}). However, if the accretion is indeed efficient,
it is likely that the accreting planets will exceed the critical luminosity
and begin experiencing eccentricity excitation due to thermal driving.
Our simulations robustly demonstrate that the eccentricities become non-zero for the majority of considered planetary masses, beginning with those as small as Mars-sized embryos.
Such eccentric planets experience inward migration because they lose the support of the corotation torque.
Planets with $M_{\mathrm{p}}\simeq4\,M_{\earth}$ undergo more complex behaviour
and they can actually remain on circular orbits. Nonetheless,
we argue that before reaching this mass,
the planet would have grown through a sequence of lower masses,
during which its eccentricity would become
excited and then maintained even upon reaching $M_{\mathrm{p}}\simeq4\,M_{\earth}$ (the planet would 
start in the headwind-dominated regime of the bottom-right panel of Fig.~\ref{fig:dedt} and it would remain there).

In Section~\ref{sec:peb_acc}, we developed
a 1D model to estimate the pebble accretion rates, which revealed that for planets with masses $M_{\mathrm{p}}\lesssim2\,M_{\earth}$, the accretion rate slows
down near the inner edge of the bump, resulting in a sub-critical luminosity (see Fig.~\ref{fig:doubl_time_map}).
However, when that occurs, the planet is already located inwards the island of outward migration (Fig.~\ref{fig:tqmap}) which is radially narrower than the radial range of super-critical luminosities.
Therefore, the planet is not saved from inward migration.

In other words, the main implication of our model is that the pressure bump is not likely
to harbour a growing embryo until it becomes a full-grown planet. Instead, 
the bump is likely to lose the embryo to inward migration. This process could be repeated
and thus the bump could spawn sub-Earth-mass embryos and populate the disc region interior
to the bump with them.

Future work is necessary to explore more parameters of the pressure bump itself, as only one pressure bump was studied here. 
For instance, one can imagine more pronounced
pressure bumps that would act as perfect barriers for drifting pebbles. In such cases, the pebble 
concentration would be confined to a narrow ring, as studied by \cite{Morbidelli_2020A&A...638A...1M}, who
showed that the range of equilibrium planet positions (without thermal torques)
would be wider than the pebble ring itself.
By incorporating thermal torques, it might be feasible
for a planet starting within the pebble ring to migrate outside
it, switch to $L<L_{\mathrm{c}}$, and still find itself
in the range of the Type-I migration trap.
However, we emphasize that any model considering pressure bumps as perfect pebble traps
should be able to account for streaming instabilities and planetesimal formation
because large solid-to-gas ratios are expected to be reached \citep[e.g.][]{Lau_etal_2022A&A...668A.170L}.
How planetesimal accretion near pressure bumps affects our study is unclear and requires further investigation, but
it can in principle only increase the energy output
of the planets.

\subsection{General implications for the assembly of planetary systems}

We would like to emphasize that even though \cite{Benitez-Llambay_etal_2015Natur.520...63B}
argued that the heating torque adds a positive contribution to the total torque
and even though the linear heating torque of \cite{Masset_2017MNRAS.472.4204M} is positive for circular orbits
in classical power-law discs,
the heating torque does not necessarily support outward migration once the eccentricity is excited
(see Fig.~\ref{fig:at_eeq}).
And since the heating torque operates at $L>L_{\mathrm{c}}$, the eccentricity excitation will inevitably
accompany it.
Therefore, current predictions for planet evolution with the addition of the heating torque, such as
those from \cite{Guilera_etal_2019MNRAS.486.5690G}, might strongly overestimate the importance
of outward migration because they do not take the eccentricity driving into account.
The inclusion of the eccentricity excitation \citep{Fromentau_Masset_2019MNRAS.485.5035F,Cornejo_etal_2023MNRAS.tmp..658C}
in N-body codes seems to be the most important missing piece at the moment because non-zero $e$
affects all components of the disc-driven torques.

If planets accrete efficiently and their eccentricities are pumped,
we think that they might ignore any migration
traps driven by positive corotation torques, even those related to the entropy-driven corotation torque \citep{Paardekooper_Mellema_2006A&A...459L..17P,Paardekooper_Mellema_2008A&A...478..245P}.
To explain the clustering and trapping of low-mass planets at the inner disc edge \citep[e.g.][]{Mulders_etal_2018AJ....156...24M,Flock_etal_2019A&A...630A.147F,Chrenko_etal_2022A&A...666A..63C},
it might be necessary for planets to switch back to sub-critical luminosities and circularize
via the usual eccentricity damping. One way to achieve this in the pebble accretion
paradigm is for the planet to reach the terminal pebble isolation mass $M_{\mathrm{iso}}$ \citep{Lambrechts_etal_2014A&A...572A..35L},
which decreases in the inner disc \citep{Bitsch_2019A&A...630A..51B}
as well as when the orbits are eccentric 
\citep{Chametla_etal_2022MNRAS.510.3867C}.
Another possibility is to rely on a decrease of the
pebble flux as the pebble disc becomes depleted \citep[e.g.][]{Appelgren_etal_2023A&A...673A.139A}
or as multiple planets contribute to the filtering of pebbles
\citep{Morbidelli_Nesvorny_2012A&A...546A..18M,Izidoro_etal_2021A&A...650A.152I}.

\section{Conclusions}
\label{sec:concl}

Previous studies have suggested that pressure bumps in protoplanetary
discs can facilitate rapid and efficient planet accretion \citep[e.g.][]{Morbidelli_2020A&A...638A...1M,Guilera_etal_2020A&A...642A.140G,Chambers_2021ApJ...914..102C,Andama_etal_2022MNRAS.512.5278A,Lau_etal_2022A&A...668A.170L,Jiang_Ormel_2023MNRAS.518.3877J}
because the Lindblad and corotation disc-driven torques
cancel out near the bump, allowing the growing planets to remain close to a reservoir of accumulating dust and pebbles.
In this study, we explored the robustness of
the migration trap when the thermal torques \citep{Lega_etal_2014MNRAS.440..683L,Benitez-Llambay_etal_2015Natur.520...63B}
are taken into account.
To this end, we conducted high-resolution 3D radiative hydrodynamic simulations, modelling the pressure bump as a Gaussian perturbation of the density and viscosity.
We focused on planets in the mass range of $M_{\mathrm{p}}=0.1$--$4\,M_{\earth}$ and
we considered that they release the accretion heat, parametrized with respect
to the critical luminosity $L_{\mathrm{c}}$ derived from the linear theory of thermal torques \citep{Masset_2017MNRAS.472.4204M}. For instance, 
Mars- and Earth-sized planets in our disc model require
the mass doubling times $\tau\simeq35$ and $170\,\mathrm{kyr}$,
respectively, to achieve $L=L_{\mathrm{c}}$.

Our study yields several key findings:
\begin{itemize}
    \item The migration trap is robust when
    the planet's luminosity is sub-critical ($L<L_{\mathrm{c}}$).
    \item For super-critical luminosities ($L>L_{\mathrm{c}}$), planets with $M_{\mathrm{p}}\lesssim2\,M_{\earth}$
    experience eccentricity excitation by thermal driving and enter the headwind-dominated regime of thermal torques \citep{Chrenko_etal_2017A&A...606A.114C,Eklund_Masset_2017MNRAS.469..206E}. 
    This excitation causes $e$ to become a sizeable fraction of the disc aspect ratio $h$ \citep[see also][]{Velasco-Romero_etal_2022MNRAS.509.5622V}, which quenches the positive corotation torque and allows the negative Lindblad torque to prevail. As a result, the planet undergoes
    orbital decay and migrates past the bump.
    Our findings suggest that although the thermal forces 
    maintain $e$ excited
    (which modifies the Lindblad and corotation torques),
    they have a negligible contribution to the migration rate in the headwind-dominated regime \citep[see also][]{Pierens_2023MNRAS.520.3286P}.
    \item For a handful of cases with super-critical luminosities ($M_{\mathrm{p}}=2\,M_{\earth}$
    with $L=3\,L_{\mathrm{c}}$, and all cases with $M_{\mathrm{p}}=4\,M_{\earth}$),
    the thermal disturbance near the planet fluctuates as in \cite{Chrenko_Lambrechts_2019}
    and the eccentricity excitation can be prevented.
    \item If the planet remains circular and evolves in the region of the bump where the gas rotation
    is super-Keplerian ($\eta<0$), the asymmetry of thermal lobes becomes reversed compared
    to the standard circular case of \cite{Benitez-Llambay_etal_2015Natur.520...63B}. The reversal
    is driven by an outward shift of the corotation between the planet and the gas.
    The inner thermal lobe leading the orbital motion deepens and the thermal
    torque becomes negative, contributing again to inward migration.
\end{itemize}
    
We also simulated a simplified 1D gas-pebble disc to estimate the mass loading of
the pressure bump by pebbles and to calculate expected pebble accretion rates.
Our results indicate that most of the planets considered in our study reach super-critical luminosities
in the vicinity of the bump. Therefore, the prevalent outcome of our model is that
growing low-mass embryos leave the bump and populate the disc interior from the bump.

Planet formation scenarios with pressure bumps should be refined by considering that
the migration trap due to the positive corotation torque is not robust in the presence
of vigorous accretion heating and eccentricity driving.
This fact can be generalized to any Type-I migration trap operating in protoplanetary discs.
Accumulation of low-mass planets at migration traps might delicately depend on the processes
regulating their accretion efficiency, such as the pebble isolation.

\section*{Acknowledgements}

This work was supported by the Czech Science Foundation (grant 21-23067M)
and the Ministry of Education, Youth and Sports of
the Czech Republic through the e-INFRA CZ (ID:90254). The
work of O.C. was supported by the Charles University Research program (No.
UNCE/SCI/023). We wish to thank the referee Elena Lega
whose valuable and constructive comments allowed us to improve this paper.

%%%%%%%%%%%%%%%%%%%%%%%%%%%%%%%%%%%%%%%%%%%%%%%%%%
\section*{Data Availability}

The public version of the Fargo3D code is available at \url{https://bitbucket.org/fargo3d/public/}. 
The simulation data underlying this article will be shared on reasonable request
to the corresponding author. The Optool code is available at
\url{https://github.com/cdominik/optool}.

%%%%%%%%%%%%%%%%%%%% REFERENCES %%%%%%%%%%%%%%%%%%

% The best way to enter references is to use BibTeX:

\bibliographystyle{mnras}
\bibliography{references} % if your bibtex file is called example.bib

%%%%%%%%%%%%%%%%%%%%%%%%%%%%%%%%%%%%%%%%%%%%%%%%%%

%%%%%%%%%%%%%%%%% APPENDICES %%%%%%%%%%%%%%%%%%%%%

\appendix

\section{Results for the half-Earth-mass planet}
\label{sec:app_halfearth}

\begin{figure}
    \centering
    \includegraphics[width=0.95\columnwidth]{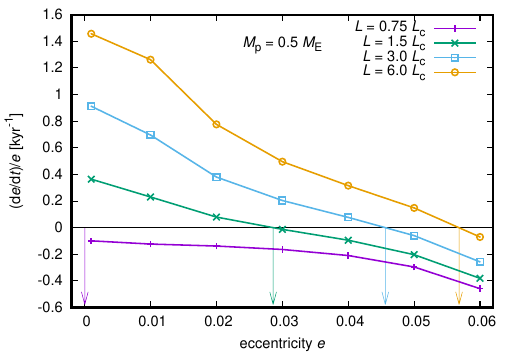}
    \includegraphics[width=0.95\columnwidth]{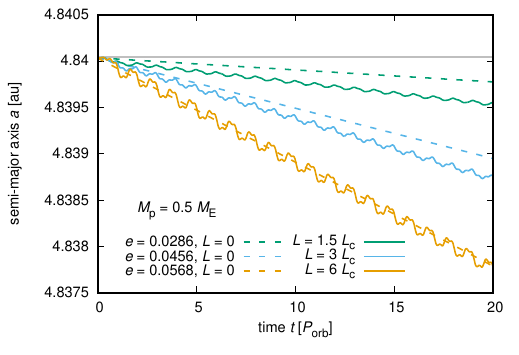}
    \includegraphics[width=0.95\columnwidth]{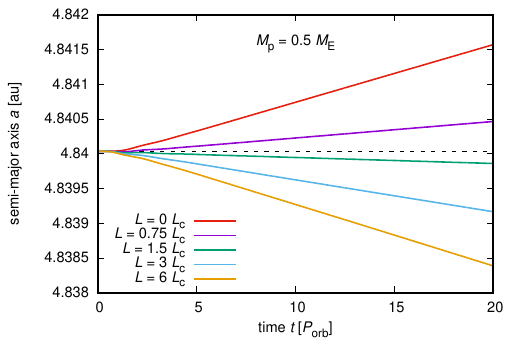}
    \caption{Supplementary figures for $M_{\mathrm{p}}=0.5\,M_{\earth}$.
    \emph{Top:} as in Fig.~\ref{fig:dedt}, \emph{middle:} as in Fig.~\ref{fig:at_eeq},
    \emph{bottom:} as in Fig.~\ref{fig:at_e0}.}
    \label{fig:append_mp0.5}
\end{figure}

With the aim to keep the main text short and
comprehensive,
this appendix summarizes simulation results
for the planetary mass $M_{\mathrm{p}}=0.5\,M_{\earth}$.
Individual panels of Fig.~\ref{fig:append_mp0.5} 
are complementary to Figs.~\ref{fig:dedt}, \ref{fig:at_eeq},
and \ref{fig:at_e0} as well as to their discussion in the main text
of the article.

\begin{figure*}
    \centering
    \begin{tabular}{cc}
    \includegraphics[width=0.95\columnwidth]{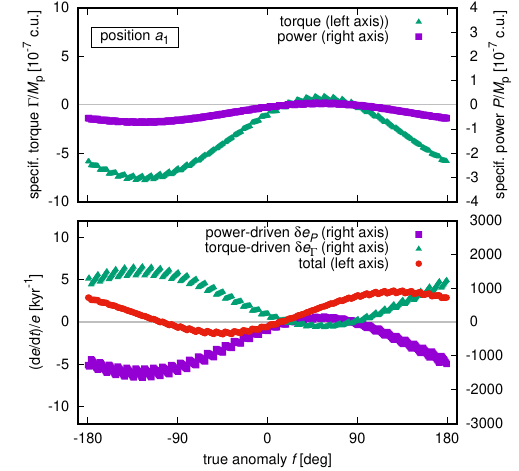} &
    \includegraphics[width=0.95\columnwidth]{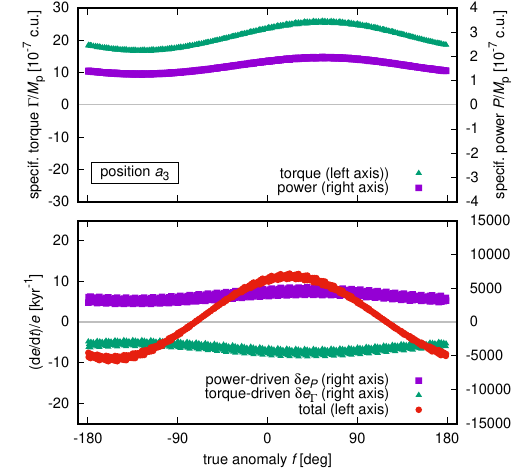} \\
    \end{tabular}
    \caption{Analysis of the torque, power, and eccentricity evolution rate for $M_{\mathrm{p}}=1\,M_{\earth}$, $L=3\,L_{\mathrm{c}}$, and $e=0.001$
    at the initial semi-major axes $a_{1}$ (\emph{left}) and $a_{3}$ (\emph{right}). \emph{Top:} Specific torque $\Gamma/M_{\mathrm{p}}$ (purple squares, primary vertical axis)
    and power $P/M_{\mathrm{p}}$ (green triangles, secondary vertical axis) as a function of the true anomaly $f$. \emph{Bottom:} Eccentricity evolution rate $\dot{e}/e$ (red circles, primary vertical axis),
    and its components driven by the disc power $\delta e_{P}$ (purple squares, secondary vertical axis)
    and disc torque $\delta e_{\Gamma}$ (green triangles, secondary vertical axis).
    Data points were recorded over the last 10 orbits of the planet.}
    \label{fig:dedt_orbital}
\end{figure*}

\section{Eccentricity growth in the shear-dominated regime}
\label{sec:app_egrowth}

We demonstrated in Section~\ref{sec:shear}, for the case of circular orbits,
that the influence of thermal torques on the semi-major axis evolution at $a_{2}$
and $a_{3}$ is in agreement with previous
studies \citep[e.g.][]{Benitez-Llambay_etal_2015Natur.520...63B},
while it can facilitate inward migration in the super-Keplerian region 
near $a_{1}$. However, the eccentricity evolution rate is quite similar
at $a_{1}$, $a_{2}$, and $a_{3}$ (Fig.~\ref{fig:dedt_acmpr}). 

To provide more insight into the eccentricity evolution in the shear-dominated
regime, we performed two simulations at $a_{1}$ and $a_{3}$ for 
the parameters $M_{\mathrm{p}}=1\,M_{\earth}$, $e=0.001$, and $L=3\,L_{\mathrm{c}}$.
These simulations were performed over 20 orbits and the planets were free to migrate.
We measured the disc torque $\Gamma$ and power $P$ with the aim to relate
them to the eccentricity evolution rate. To do so, one can utilize \citep{Bitsch_Kley_2010A&A...523A..30B}
\begin{equation}
    \frac{\dot{e}}{e} = \frac{1-e^{2}}{e^{2}}\left(\frac{1}{2}\frac{\dot{a}}{a}-\frac{\Gamma}{L}\right)
    \coloneqq \delta e_{P} + \delta e_{\Gamma} \, ,
    \label{eq:dedt}
\end{equation}
where $\dot{a}/a$ comes from Equation~(\ref{eq:dadt}) and the orbital angular momentum is
\begin{equation}
    L = \mu\sqrt{GMa\left(1-e^{2}\right)} \, ,
    \label{eq:angmom}
\end{equation}
where $M=M_{\star}+M_{\mathrm{p}}$ and $\mu=M_{\star}M_{\mathrm{p}}/M$. In writing Equation~(\ref{eq:dedt}),
we split the expression into a power-driven term $\delta e_{P}$ and a torque-driven term $\delta e_{\Gamma}$.

Fig.~\ref{fig:dedt_orbital} shows $\Gamma$, $P$, $\delta e_{\Gamma}$, $\delta e_{P}$, and $\dot{e}/e$
as functions of the true anomaly $f$. We can therefore infer how the instantaneous change of the orbital 
angular momentum and energy propagates into the change of $e$ during orbital cycles of the planet.
First, we notice that $P$ is negative at $a_{1}$ while it is positive at $a_{3}$. This corresponds
to the opposite $\dot{a}$ at these two locations (see Equation~\ref{eq:dadt}). 
Second, $\dot{e}$ oscillates during each orbit but the positive values clearly
dominate on average and as a result, $e$ grows both at $a_{1}$ and $a_{3}$.
However, there is a systematic difference between the profiles of $\dot{e}(f)/e$. For $a_{1}$,
$\dot{e}$ is zero when the planet is at the pericentre, then it follows a broad positive peak
as the planet moves towards the apocentre, and a small negative peak occurs between $f=-90^{\circ}$ and $0^{\circ}$.
For $a_{3}$, the maximum (minimum) of $\dot{e}$ occurs roughly at the pericentre (apocentre).
%Finally, it is interesting to note
%that the terms $\delta e_{P}$ and $\delta e_{\Gamma}$ are by orders of magnitude
%larger than their actual sum.

\section{Connection to detailed opacity models}
\label{sec:opa}

In our disc model, we assumed
a constant uniform opacity $\kappa=1\,\mathrm{cm}^{2}\,\mathrm{g}^{-1}$.
Whether this value is realistic or not depends mostly on the dust composition
and grain sizes in the given disc region. To provide at least one quantitative comparison
of our $\kappa$ with a more detailed opacity model, we calculated frequency-dependent
dust opacities $\kappa_{\nu,\mathrm{d}}$ using the Optool code \citep{Dominik_OPTOOL_2021ascl.soft04010D},
while assuming the DIANA standard for the dust grain composition and size-frequency distribution \citep{Woitke_etal_2016A&A...586A.103W}.
Subsequently, we converted $\kappa_{\nu,\mathrm{d}}$ into the Rosseland mean opacities $\kappa_{\mathrm{R}}$,
which are commonly used to describe the transfer of thermal radiation in the gray diffusion approximation.
Finally, since $\kappa$ is used in our model to calculate optical depths in the gas, 
we rescaled $\kappa_{\mathrm{R}}$ by multiplying it with the canonical dust-to-gas ratio 
of the interstellar medium $f_{\mathrm{d2g}}=0.01$.

\begin{figure}
    \centering
    \includegraphics[width=0.95\columnwidth]{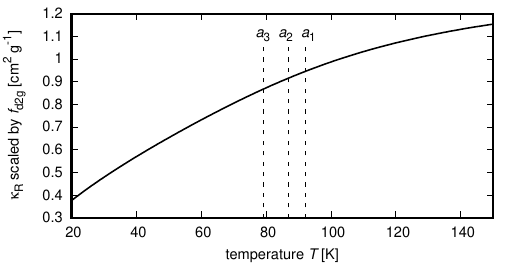}
    \caption{Frequency-averaged Rosseland opacity $\kappa_{R}$
    as a function of the local disc temperature $T$. To calculate $\kappa_{R}$,
    we considered the standard DIANA dust grain composition and 
    we also scaled $\kappa_{R}$ by the typical dust-to-gas ratio $f_{\mathrm{d2g}}=0.01$.
    The vertical dashed lines show the temperature of our unperturbed disc at locations
    $a_{1}$, $a_{2}$, and $a_{3}$. The opacity curve can be readily compared with
    the choice of the uniform opacity $\kappa=1\,\mathrm{cm^{2}}\,\mathrm{g}^{-1}$
    used in our simulations.}
    \label{fig:opa}
\end{figure}

Fig.~\ref{fig:opa} shows the scaled Rosseland opacities as a function of the local
disc temperature $T$. At radial positions $a_{1}$, $a_{2}$, and $a_{3}$, which are highlighted
with vertical dashed lines, we can see that $\kappa=1\,\mathrm{cm}^{2}\,\mathrm{g}^{-1}$
differs from the Rosseland opacity curve by 6, 9, and 15 per cent, respectively.

%%%%%%%%%%%%%%%%%%%%%%%%%%%%%%%%%%%%%%%%%%%%%%%%%%

% Don't change these lines
\bsp	% typesetting comment
\label{lastpage}
\end{document}